\documentclass[a4paper,11pt]{article}
\pdfoutput=1
\usepackage[utf8]{inputenc}
\usepackage[T1]{fontenc}
\usepackage[numbers, compress, elide]{natbib}
\usepackage[no-natbib-sort]{jheppub}

\usepackage[usenames,dvipsnames,svgnames,table,x11names]{xcolor}
\usepackage{lmodern,amsmath,multirow,cancel}
\usepackage{booktabs}
\usepackage{xstring}
\usepackage{ifthen}

\usepackage{ascmac,braket,bm,mathrsfs,amsthm,amsfonts}
\usepackage{hyperref}
\usepackage{graphicx}
\usepackage{subcaption} 
\usepackage{comment}

\usepackage{titlesec}
\newcommand*{\justifyheading}{\raggedright}
\titleformat{\chapter}[display]
  {\normalfont\huge\bfseries\justifyheading}{\chaptertitlename\ \thechapter}
  {20pt}{\Huge}
\titleformat{\section}
  {\normalfont\Large\bfseries\justifyheading}{\thesection}{1em}{}
\titleformat{\subsection}
  {\normalfont\large\bfseries\justifyheading}{\thesubsection}{1em}{}
\titleformat{\subsubsection}
  {\normalfont\bfseries\justifyheading}{\thesubsubsection}{1em}{}
\usepackage[english]{babel}
\usepackage[dvipsnames]{xcolor}
\numberwithin{equation}{section}

\renewcommand{\star}{\ast}

\usepackage{slashed}
\newcommand{\Slash}[1]{{\ooalign{\hfil/\hfil\crcr$#1$}}}

\renewcommand\Re{\mathop{\mathrm{Re}}}
\renewcommand\Im{\mathop{\mathrm{Im}}}

\newcommand{\ie}{{\em i.e.}}

\newcommand{\sla}[1]{\slashed{#1}}
\newcommand{\gsim}{ \mathop{}_{\textstyle \sim}^{\textstyle >} }
\newcommand{\lsim}{ \mathop{}_{\textstyle \sim}^{\textstyle <} }

\preprint{IPMU23-0002,~KEK-TH-2494}

\title{Novel loop-diagrammatic approach 
to QCD \boldmath{$\theta$} parameter and application to the left-right model}

\author[a,b,c]{Junji Hisano,}
\author[b,d,e,f]{Teppei Kitahara,}
\author[a]{Naohiro Osamura,}
\author[a]{and Atsuyuki Yamada}

\affiliation[a]{
  Department of Physics, Nagoya University, Furo-cho Chikusa-ku, Nagoya 464-8602 Japan
}
\affiliation[b]{
  Kobayashi-Maskawa Institute for the Origin of Particles and the
  Universe, Nagoya University,
  Furo-cho Chikusa-ku, Nagoya 464-8602 Japan
}
\affiliation[c]{
  Kavli IPMU (WPI), UTIAS, The University of Tokyo, Kashiwa 277-8584, Japan
}
\affiliation[d]{
Institute for Advanced Research, Nagoya University, Furo-cho Chikusa-ku, Nagoya 464-8601, Japan}
\affiliation[e]{
KEK Theory Center, IPNS, KEK, Tsukuba 305--0801, Japan}
\affiliation[f]{
  CAS Key Laboratory of Theoretical Physics, Institute of Theoretical Physics, Chinese Academy of Sciences, Beijing 100190, China}

\emailAdd{hisano@eken.phys.nagoya-u.ac.jp}
\emailAdd{teppeik@kmi.nagoya-u.ac.jp}
\emailAdd{osamura.naohiro.j2@s.mail.nagoya-u.ac.jp}
\emailAdd{yamada.atsuyuki.k3@s.mail.nagoya-u.ac.jp}

\abstract{
When the QCD axion is absent in full theory, 
the strong $CP$ problem has to be explained by an additional mechanism, {\it e.g.}, the left-right symmetry. 
Even though tree-level QCD $\bar\theta$ parameter is restricted by the mechanism, radiative corrections to $\bar\theta$ are mostly generated, which leads to a dangerous neutron  electric dipole moment (EDM). 
The ordinary method for calculating the radiative $\bar\theta$
utilizes an equation $\bar \theta= -\text{arg}\, \text{det}\, m_q^{\rm loop}$
 based on the chiral rotations of complex quark masses.
 In this paper, we point out that
 when full theory includes extra heavy quarks,
 the ordinary method is unsettled for the extra quark contributions and does not contain its full radiative corrections.
 We formulate a novel method to calculate the radiative corrections to $\bar\theta$ through a direct loop-diagrammatic approach, which should be more robust than the ordinary one.
 As an application,
we investigate the radiative $\bar\theta$  
in the minimal left-right symmetric model.
We first confirm a seminal result that 
two-loop level radiative $\bar\theta$ completely vanishes 
(corresponding to one-loop corrections to the quark mass matrices).
Furthermore, 
we estimate the size of a non-vanishing radiative $\bar\theta$  at three-loop level.
It is found that the resultant induced neutron EDM is comparable to the current experimental bound, 
and the expected size is restricted by the perturbative unitarity bound 
in the minimal left-right model. 
}


\begin{document}
\sloppy 

\makeatletter\renewcommand{\@fpheader}{\ }\makeatother

\maketitle

\renewcommand{\thefootnote}{\#\arabic{footnote}}
\setcounter{footnote}{0}

\section{Introduction}

The QCD $\theta$ term 
is $P$- and $T$-odd, and then $CP$-odd under the $CPT$ invariance. 
Because it is identical to the total derivative, 
it never locally affects physics at the classical level (as long as the momentum conservation holds), while
its effect occurs only via nonperturbative processes  \cite{Belavin:1975fg,tHooft:1976rip,Callan:1976je,Jackiw:1976pf,Witten:1979ey}.
It is known that this interaction 
induces the neutron electric dipole moment (EDM) \cite{Baluni:1978rf,Crewther:1979pi}.
Measurement of the neutron EDM by the nEDM collaboration has set the severe upper bound: 
$|d_n|_{\rm exp} < 1.8 \times  10^{-26} ~e\,\text{cm}$ (90\% CL)~\cite{Abel:2020pzs}.
%
Using the latest lattice result $d_n = - 0.00148(34)\, \bar{\theta}\, e\,\text{fm}$
\cite{Liang:2023jfj}\footnote{%
The first nonzero calculation by using the lattice QCD simulation was achieved in Ref.~\cite{Dragos:2019oxn}, 
then the first statistically significant result was obtained in Ref.~\cite{Liang:2023jfj}.}
and assuming that $\bar{\theta}$ is the only source of $CP$ violation,
one obtains the upper bound on the angle, 
\begin{eqnarray}
  |\bar{\theta} | \lesssim 1.2 \times 10^{-10} \quad (90\%~\text{CL})\,,
  \label{eq:constraint theta}
\end{eqnarray}
where $\bar\theta$ is the physical $CP$-violating angle in the QCD Lagrangian which will be 
 defined explicitly in the next section.

Although the experimental bound requires that $\bar{\theta}$ must be around zero, 
such a $CP$-violating phase is not restricted in 
the Standard Model (SM). 
In fact, the $CP$-violating phase in the Cabibbo-Kobayashi-Maskawa
(CKM) matrix \cite{Cabibbo:1963yz,Kobayashi:1973fv} 
is $\mathcal{O}(1)$; $\delta_{\rm CKM}\simeq 66^\circ = 1.2$\,rad \cite{Charles:2004jd} (in the standard parameterization  \cite{Chau:1984fp,ParticleDataGroup:2022pth}). 
If there is no trick in the full theory, 
$\bar\theta \ll 1$, or equivalently $\bar{\theta} \ll \delta_{\rm CKM}=\mathcal{O}(1)$, requires a fine-tuning 
at $\mathcal{O}(10^{-10})$ level.
This is known as the strong $CP$ problem.

The massless up quark could be a solution to the strong $CP$ problem if the observed hadron masses are explained by the nonperturbative effect \cite{Georgi:1981be,Kaplan:1986ru,Choi:1988sy,Banks:1994yg,Srednicki:2005wc}. However, 
some lattice studies ruled out this solution \cite{Alexandrou:2020bkd}.
Thus, the strong $CP$ problem would suggest that the SM has to be extended to suppress $\bar\theta$ without the fine-tuning. Axion is the simplest solution to the strong $CP$ problem \cite{Peccei:1977hh, Weinberg:1977ma, Wilczek:1977pj}, though it suffers from another fine-tuning in the quantum gravity sector (axion quality problem) \cite{Kamionkowski:1992mf,Holman:1992us,Barr:1992qq}.

Alternatively, one may resolve the strong $CP$ problem by extended parity symmetry \cite{Beg:1978mt,Mohapatra:1978fy,Babu:1989rb, Barr:1991qx}.\footnote{%
Other possibilities are spontaneous $CP$ violation referred to as the Nelson-Barr mechanism \cite{Nelson:1983zb,Barr:1984qx,Barr:1984fh}.}
In such scenarios, the extended parity involves the left-right (LR) gauge symmetry.
The parity symmetry forbids the bare $\bar\theta$ parameter, while $\mathcal{O}(1)$ of $\delta_{\rm CKM}$ is allowed.
It is known that even though the bare $\bar\theta$ parameter is strictly forbidden by the parity symmetry, radiative correction to $\bar\theta$ is regenerated since the parity symmetry must be softly broken in nature. Eventually, one has to consider the experimental bound on the model from the neutron EDM measurements in Eq.~\eqref{eq:constraint theta} through the radiatively regenerated $\bar\theta$.

The ordinary method for calculating the radiatively generated $\bar\theta$ parameter, adopted in many papers, utilizes 
\begin{align}
\bar \theta= -\text{arg}\, \text{det} \, m_u^{\rm loop} - \text{arg}\, \text{det} \, m_d^{\rm loop}\,.
\end{align}
Here, $m_{u,d}^{\rm loop}$ are the up- and down-type quark mass matrices including the radiative corrections. This relation is
 based on the chiral rotations for the complex quark masses  and an anomalous divergence of the axial-vector current, known as the Adler-Bell-Jackiw anomaly \cite{Adler:1969gk,Bell:1969ts}.
Or,  it is also derived using the path-integral formalism referred to as the Fujikawa method~\cite{Fujikawa:1979ay}.

The ordinary method is simple though it may not be accurate.  For instance, within the SM, even if one sets the bare $\bar{\theta}$ parameter to be zero, it is radiatively produced via the $CP$-violating phase in the CKM matrix. It is shown with the above method in Ref.~\cite{Ellis:1978hq} that the contribution via the radiative corrections to quark masses is of $\mathcal{O}(G_F^2\alpha_s^3)$ ($G_F$ is the Fermi coupling constant and $\alpha_s$ is the QCD coupling constant). On the other hand, the direct loop calculation of the correction to $\bar\theta$ shows that it is derived at four-loop order  ($\mathcal{O}(G_F^2\alpha_s)$) \cite{Khriplovich:1985jr}.\footnote{%
A hadronic long-distance evaluation has confirmed that the radiative $\bar{\theta}$ parameter is produced at order of $\mathcal{O}(G_F^2\alpha_s)$ \cite{Gerard:2012ud}.} It is consistent with the fact that the EDMs (and also the chromo-EDMs) of quarks are induced at three-loop order ($\mathcal{O}(G_F^2\alpha_s)$) \cite{Czarnecki:1997bu}.
In fact, the ordinary method corresponds to the diagrams where the external gluons are attached in the same fermion line in loop diagrams contributing to $\bar\theta$. 
It implies that
the leading $\bar\theta$ in the SM \cite{Khriplovich:1985jr} comes from diagrams with external gluons attached to different fermion lines.

In this paper, we formulate a novel approach to evaluate the radiative corrections to $\bar\theta$ through a direct loop-diagrammatic calculation,  which should be more robust than the ordinary one. In Ref.~\cite{Khriplovich:1985jr}, the external gluon field is introduced to calculate the correction to $\bar\theta$ from the CKM matrix,  while details of the technique are not written. We introduce the Fock-Swinger gauge method to directly calculate the radiative corrections to  $\bar\theta$ under the gluon field-strength background. 

As an application, we investigate the radiative $\bar{\theta}$ in the minimal LR symmetric model 
\cite{Babu:1989rb,Barr:1991qx}, in which the bare and one-loop level $\bar\theta$ parameters are strictly forbidden by the LR symmetry. Although the extra heavy quarks  whose Yukawa interactions violate $CP$ symmetry are introduced, 
the $CP$-violating Yukawa interactions do not contribute to the  $\bar\theta$ parameters at one-loop level. 
Furthermore, it is known that two-loop level  $\bar\theta$ parameter also vanishes, 
 which corresponds to one-loop corrections to the quark masses in the ordinary method 
 \cite{Babu:1989rb, Hall:2018let, Craig:2020bnv}.  However, the ordinary method is unsettled for the extra quark contributions to $\bar\theta$ and indeed does not contain its full radiative corrections, like the SM calculations \cite{Khriplovich:1985jr}.
We first confirm this seminal result by using the proposed  
method. While new type diagrams contribute to $\bar\theta$ at two-loop level, 
the sum of the diagrams still gives no contribution to $\bar\theta$. Next, we estimate the size of a non-vanishing
radiative $\bar\theta$  at three-loop level. It will be found that the resultant induced neutron EDM is 
comparable to the current experimental bound. We will also investigate a relation between the radiative three-loop
level $\bar\theta$ and 
the perturbative unitarity bounds of the Yukawa couplings in the minimal LR symmetric model.

This paper is organized as follows.
In Sec.~\ref{sec:method}, we discuss methods of direct calculation of the loop diagrams contributing to $\bar\theta$. We show that the operator Schwinger method and the Fock-Schwinger gauge method are applicable, though 
the latter method has a merit to extend the calculation to the higher-loop diagrams. 
In Sec.~\ref{sec:model}, the minimal LR symmetric model is briefly summarized. 
We also derive the parameterization by the physical
parameters based on the seesaw mechanism in the LR symmetric model.
We confirm that the two-loop level radiative $\bar\theta$ vanishes by using the proposed method  in Sec.~\ref{sec:proof}.
In Sec.~\ref{sec:3loop}, we investigate the numerical size of the non-vanishing radiative $\bar\theta$ and compare both experimental (from neutron EDM) and theoretical bounds (from the perturbative unitarity bound).
Section~\ref{sec:conclusions} is devoted to conclusions and discussion.
Details of the loop calculations are given in the Appendix.

\section{Loop-diagrammatic evaluation of QCD \texorpdfstring{\boldmath{$\theta$}}{theta} parameter}
\label{sec:method}

In the QCD Lagrangian,  imaginary parts of the quark masses and the QCD $\theta$ term are $P$-odd and $T$-odd interactions that are not restricted from the $SU(3)_C$ gauge symmetry, 
\begin{align}
    \mathcal{L}_{\Slash{P},\,\Slash{T}}
    = - \sum_{q=\text{all}} \text{Im}(m_q) \bar{q} i \gamma_5 q + \theta_G \frac{\alpha_s}{8\pi } G_{\mu\nu}^a\tilde{G}^{a\mu\nu}\,,
\end{align}
where $m_q$ stands for the complex quark masses with $m_q \equiv |m_q| \exp(i \theta_q)$, 
$G_{\mu\nu}^a$ is the gluon field-strength tensor, 
$\tilde{G}^{a\mu\nu} \equiv \frac{1}{2} \epsilon^{\mu \nu \rho \sigma} G^a_{\rho \sigma}$ with $\epsilon^{0123} = +1$, 
and $\alpha_s =g_s^2/(4\pi)$ is the $SU(3)_C$ coupling constant.
It is well-known that the axial rotation of quarks 
\begin{align}
q \to q^\prime = \exp {\left(- \frac{i}{2}\theta_q \gamma_5\right)}\, q \,,
\qquad 
\bar{q} \to \bar{q}^\prime = \bar{q} \exp {\left(- \frac{i}{2}\theta_q \gamma_5\right)}\,,
\end{align}
turns off the imaginary part of the quark masses and 
generates an additional QCD $\theta$ term,
\begin{align}
    \mathcal{L}_{\Slash{P},\,\Slash{T}}
    =  
{\bar{\theta}} \, \frac{\alpha_s}{8\pi }  G_{\mu\nu}^a\tilde{G}^{a\mu\nu}\,,
    \label{eq:chiral_rotation}
\end{align}
where 
\begin{align}
\bar{\theta}\equiv \theta_G -  \sum_q\theta_q\,,
\label{phsical_theta}
\end{align}
is a physical $\bar\theta$ parameter,  if all quarks are massive. This is derived  using the path-integral formalism referred to as the Fujikawa method~\cite{Fujikawa:1979ay} or with the Adler-Bell-Jackiw anomaly \cite{Adler:1969gk,Bell:1969ts,Adler:1969er,tHooft:1972tcz}
\begin{align}
\partial_{\mu} (\bar{q} \gamma^{\mu} \gamma_5 q)
=  2 i \Re (m_q) \bar{q} \gamma_5 q - 2 \Im (m_q)\bar{q} q - \frac{\alpha_s}{ 4 \pi} G^a_{\mu\nu} \tilde G^{ a \mu \nu}\,.
 \end{align}

The contribution of the quark mass phase to the QCD $\theta$ term should be able to be directly evaluated by loop-diagrammatically integrating out quarks, not via the Adler-Bell-Jackiw anomaly or transformation of measure  in the path integral (the Fujikawa method).
However, it does not generate the QCD $\theta$ term
if the momenta in the diagrams are conserved.
It is because
the $\theta$ term is equivalent to
total-derivative and the total momentum has to be zero.
Thus, we have to abandon the momentum conservation  or equivalently the translation invariance in order to evaluate the QCD $\theta$ term with the loop-diagrammatic calculation. 

It can be realized by introducing the gluon field strength background. In this section, we evaluate the QCD $\theta$ term with two different methods, 1) the operator Schwinger method and 2) the Fock-Schwinger gauge method. We show that they produce  consistent results with Eq.~(\ref{phsical_theta}).\footnote{%
An alternative way to evaluate the QCD $\theta$ term is the $CP$-odd
spurion trick \cite{Georgi:1980cn}. Equation (\ref{phsical_theta}) can be derived by introducing a spurion, whose vacuum expectation value produces the $CP$-violating phase of the quark mass, with an external momentum injection via the spurion.} 

\subsection{Operator Schwinger method} 

First, we consider the operator Schwinger method.\footnote{See Ref.~\cite{Novikov:1983gd} for the review about the operator Schwinger method.}  The effective action $\Delta S$ induced by the integration of quarks at one-loop level is given by the log-determinant (or trace-log) of the Dirac operator. Now we introduce the complex mass parameters for quarks. In this case, the effective action is given as 
\begin{eqnarray}
\Delta S &=&-i {\rm Tr} \,\log \left|\left|\Slash{P}-(m_q^\ast P_L+m_q  P_R) \right|\right|\,,
\end{eqnarray}
where $P_\mu=iD_\mu$,
$D_\mu\equiv \partial_\mu+i g_s T^a G_\mu^a$
is the QCD covariant derivative, and $P_{L/R}=(1\mp\gamma_5)/2$. 
In this method, the following basic commutation relation is used,
\begin{eqnarray}
[{P}_\mu, {P}_\nu] &=& - ig_s T^a  G^a_{\mu\nu}(\equiv -ig_s G_{\mu\nu})\,,
\end{eqnarray}
since the gluon field-strength tensor appears from it. 
Then, the derivative of $\Delta S$ over the fermion mass $m_q$ for $P_R$ is 
\begin{eqnarray}
\frac{d}{dm_q}\Delta S&=&i {\rm Tr}\left[\frac{1}{\Slash{{P}}-(m_q^\ast P_L+m_q P_R)} P_R\right]\nonumber\\
&=& i  {\rm Tr}\left[\frac{1}{{P}^2-|m_q|^2 -\frac{1}{2}g_s\sigma_{\mu\nu}G^{\mu\nu}}
m_q^\star P_R\right]\,,
\end{eqnarray}
where $\sigma_{\mu\nu}= \frac{i}{2}[\gamma_{\mu},\gamma_{\nu}]$.
Since the Levi-Civita tensor appears from the trace of a product of four $\gamma$'s  and $\gamma_5$,
the second order of $\frac{1}{2} g_s\sigma_{\mu\nu}G^{\mu\nu}$ leads to the $G\tilde{G}$ term as
\begin{align}
 \frac{d}{dm_q}\Delta S \supset & i  {\rm Tr}\left[\frac{1}{{P}^2-|m_q|^2}\left(\frac{1}{2} g_s\sigma_{\mu\nu}G^{\mu\nu}\right)
\frac{1}{{P}^2-|m_q|^2}\left(\frac{1 }{2} g_s \sigma_{\rho\sigma}G^{\rho\sigma}\right)\frac{1}{{P}^2-|m_q|^2}
m_q^\star P_R\right] \nonumber\\
= & 
\int d^4x \int\frac{d^4p}{(2\pi)^4}
\frac{ g_s^2}{2} \frac{-m_q^\star}{(p^2-|m_q|^2)^3} G_{\mu\nu}^a\tilde{G}^{a\mu\nu} + \cdots 
\nonumber \\
\supset &
\int d^4x \frac{i}{32\pi^2} \frac{ g_s^2}{2} \frac1{m_q} G_{\mu\nu}^a\tilde{G}^{a\mu\nu}
\,,\label{osm}
\end{align}
where 
$\text{Tr}(T^aT^b)=(1/2)\delta^{ab}$ is used.
Here, $P^2$ is replaced by $-\partial^2$, and it is integrated in momentum space.  Similarly, 
$d\Delta S/dm_q^\star$ leads to the ${G}\tilde{G}$ term. By Integrating $d\Delta S/dm_q$ with $m_q$ and $d\Delta S/dm_q^\star$ with $m_q^\star$, we get
\begin{align}
\Delta \mathcal{L} = & \frac{-i}{32\pi^2}\frac{ g_s^2}{2} \log\frac{m_q^\star}{m_q}
G_{\mu\nu}^a\tilde{G}^{a\mu\nu} \nonumber \\
=&
- {\theta_q} \frac{\alpha_s}{8\pi} 
G_{\mu\nu}^a\tilde{G}^{a\mu\nu}\,.
\label{eq:Shwinger_method}
\end{align}    
Now we obtain the contribution from a quark with complex mass to the QCD $\theta$ term by integrating out the quark, which is consistent with the axial rotation \eqref{eq:chiral_rotation}.

\subsection{Fock-Schwinger gauge method} 
\label{sec:FSG}
    In the previous section, we showed that the operator Schwinger method enables us to derive the physical QCD $\theta$ term by the diagrammatic evaluation. However, the operator Schwinger method  is not suitable to calculate effective operators induced at higher loops because it is the method to obtain an effective action by integrating fermions out with the log-determinant of the Dirac operator. 
    Here alternatively, we introduce the Fock-Schwinger gauge method, which is more applicable to diagrammatic calculation.\footnote{See Ref.~\cite{Novikov:1983gd} for the review about the Fock-Schwinger gauge method.} 
    
    The Fock-Schwinger gauge is to take such a gauge 
    \begin{align}
        (x^\mu - x^\mu_0) G_\mu^a (x) = 0\,,
    \end{align} 
    which violates the translation symmetry. 
    Because of breaking the translation symmetry, we can derive perturbatively the QCD $\theta$ term as will be shown below. While the Fock-Schwinger gauge violates the translation symmetry, the physical observables do not depend on it. 
    In the below argument, we take gauge dependence parameter $x_0$ as $x_0=0$ for simplicity. 
    
    The gluon field $G_\mu^a$ can be expanded under this gauge around $x=0$ and it is given with the gluon field-strength tensor at $x=0$, $G^a_{\mu \nu} (0)$, as \cite{Novikov:1983gd}
    \begin{align}
            G_\mu^a (x) &= \frac{1}{2} x^\nu G_{\nu \mu}^a (0) + \cdots \nonumber \\
            &= \int d^4 k e^{-i k \cdot x} \left( - \frac{i}{2} G^a_{\nu \mu} (0) \frac{\partial}{\partial k_\nu} \delta^{(4)} (k) \right) + \cdots \,.
        \label{eq:Fock-Schwinger gauge}
    \end{align}
Here, the discarded terms are covariant derivatives of the background gluon field-strength tensor, which are irrelevant to the calculation of the QCD $\theta$ term.  We can systematically evaluate the interaction of the propagating quarks with the background gluon field-strength tensor in this gauge fixing. However, 
we found that the effective gluon operators such as the QCD $\theta$ term  cannot be evaluated from the simple quark bubble diagrams. The background gluon fields bring momenta, $k$ in Eq.~(\ref{eq:Fock-Schwinger gauge}), which are taken to be zero in the last step of the calculation due to $\delta^{(4)} (k)$. Thus, the quark momentum is not constant due to interaction with the background field and the quark line cannot be closed without violating momentum conservation.

    \begin{figure}[t]
        \centering
        \includegraphics[width=0.5\linewidth]{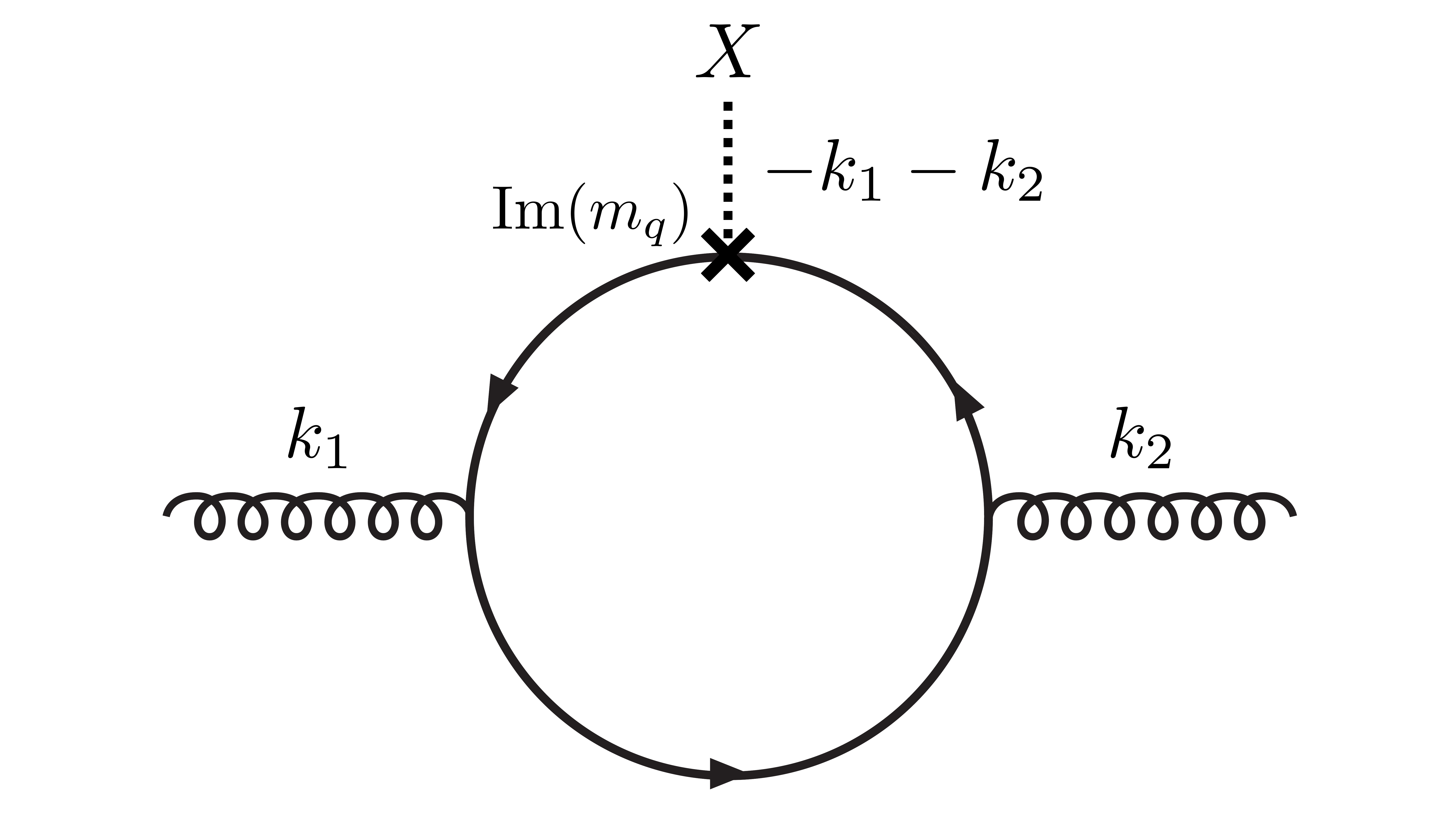}
        \caption{Feynman amplitude $i \Pi_X^q$ for the loop-diagrammatic evaluation of the $\bar\theta$ parameter.}
        \label{fig:diagrammatic evaluation}
    \end{figure}
In order to fix this problem, we introduce an auxiliary (dimensionless) background field $X$, and it is coupled to the $CP$-odd quark mass terms as 
    \begin{align}
        \mathcal{L}_{\Slash{P},\,\Slash{T}} &= - \sum_{q=\text{all}} \text{Im}(m_q) \bar{q} i \gamma_5 q \nonumber \\
        &\Rightarrow - \sum_{q=\text{all}}\text{Im}(m_q)\left( \bar{q} i \gamma_5 q \right) X\,.
        \label{eq:modified $CP$-odd mass}
    \end{align}
We evaluate the leading contribution of $\text{Im}(m_q)
$ to the QCD $\theta$ term in perturbative way, assuming $\text{Im}(m_q)\ll \text{Re}(m_q)$. 
The field $X$ is taken to be $1$ in
the last step of the calculation.\footnote{This technique has been applied for evaluation of the Weinberg operator in  the QCD~\cite{Abe:2017sam}.} 

The radiative QCD $\theta$ term comes from a bubble diagram.
The Feynman diagram in Fig.~\ref{fig:diagrammatic evaluation} shows the leading contribution, which is realized by integrating the delta function in Eq.~(\ref{eq:Fock-Schwinger gauge}) as 
    \begin{equation}
        \begin{split}
            &i \Pi_X^q = - \int d^4 k_1d^4 k_2  \frac{d^4 p}{(2 \pi)^4} \tilde{X} (-k_1-k_2) \\
            &\quad\times {\rm Tr} \left[ (-i g_s \gamma^\mu T^a) \left( - \frac{i}{2} G^a_{\rho \mu} (0) \frac{\partial}{\partial k_{1;\rho}} \delta^{(4)} (k_1)\right) \frac{i}{\sla{p}+\sla{k}_1 - {\rm Re} (m_q)}  {\rm Im} (m_q)  \gamma_5   \right. \\
            &\quad\times \left. \frac{i}{\sla{p}-\sla{k}_2-{\rm Re} (m_q)} (-i g_s \gamma^\nu T^b) \left(- \frac{i }{2} G^b_{\sigma \nu} (0) \frac{\partial}{\partial k_{2;\sigma}} \delta^{(4)} (k_2)\right) \frac{i}{\sla{p}-{\rm Re} (m_q)} \right] \,,
        \end{split}
        \label{eq:GGtilde Fock-Schwinger gauge}
    \end{equation}
    where $\tilde{X} (k) = \int d^4 x e^{i k \cdot x} X (x)$ and $p$ is the loop momentum.  We followed the Feynman rules under the Fock-Schwinger gauge, which includes the gluon field $G^a_\mu$ expressed as Eq.~(\ref{eq:Fock-Schwinger gauge}) and the modified $CP$-odd quark mass term in Eq.~(\ref{eq:modified $CP$-odd mass}).
    Until integration of the delta functions, two independent momenta $k_1, k_2$ flow into the vertices with the background field-strength tensors
    $G_{\rho \mu}^a$ and $G_{\sigma \nu}^b$, respectively, and the artificial background field $X$ brings a momentum $- k_1 - k_2$ (see Fig.~\ref{fig:diagrammatic evaluation}). After some calculation, we get
    \begin{equation}
        \begin{split}
            i \Pi_X^q &
            = i \, {\rm Im} (m_q) \frac{g_s^2}{8} G^a_{\rho \mu} (0) G^a_{\sigma \nu} (0)  \int \frac{d^4 p}{(2 \pi)^4} \frac{4 i {\rm Re} (m_q) \epsilon^{\mu \nu \rho \sigma}}{[p^2 - ({\rm Re} (m_q))^2]^3} \tilde{X} (0) \\
            &= i \frac{\alpha_s}{8 \pi} \left( - \frac{{\rm Im} (m_q)}{{\rm Re} (m_q)} \right) G^a_{\mu \nu} (0) \tilde{G}^{a \mu \nu} (0) \\
             & \simeq - i \theta_q  \frac{\alpha_s}{8 \pi}   G^a_{\mu \nu} (0) \tilde{G}^{a \mu \nu} (0)\,.
        \end{split}
        \label{FGgauge_oneloop}
    \end{equation}
Here, we take $\tilde{X}(0)=1$.

After integrating the quark $q$ out in the full theory, $\Delta \mathcal{L} = \Pi_X^q$ is obtained
in the effective action of the gluon.
 Eventually, one can derive the QCD $\theta$ term in the Fock-Schwinger gauge method. 
    This result is consistent with that of the chiral rotation, $q \to q' = 
    \exp(- \frac{i}{2} \theta_q \gamma_5 ) \,
    q$, in Eq.~(\ref{eq:chiral_rotation}) 
    and also the operator Schwinger method in Eq.~\eqref{eq:Shwinger_method}.
    Hence, we reached a clarification of the equivalence among the Fock-Schwinger gauge method, the operator Schwinger method, and the ordinary chiral rotation, and we noticed that the Fock-Schwinger gauge method is more intuitive than the operator Schwinger method for higher-loop order calculations.

It might be concerned that the  diagrammatic evaluations of 
the light quark contribution to the QCD $\theta$ term is not justified from the viewpoint of perturbation, since the loop momentum around the quark mass dominates the integrals in Eqs.~(\ref{osm}) and (\ref{FGgauge_oneloop}). It might be healthy to evaluate the light quark mass phases above the $\Lambda_{\rm QCD}$ scale and derive the QCD $\theta$ parameter by the chiral rotation. However, since the  diagrammatic evaluations are consistent with those of the  chiral rotation, we may forget the problem in practical cases.

    In this paper, to evaluate the QCD $\theta$ term diagrammatically, 
    we will use the Fock-Schwinger gauge method.
    Note that the auxiliary background field $X$ should be attached to any perturbative interactions, but we suppress them in the following calculations for the sake of clarity.

\section{The minimal left-right symmetric model}
\label{sec:model}
\subsection{Model}
    From this section, we introduce 
    the minimal LR symmetric model that can solve the strong $CP$ problem.
    The LR symmetry,
which is formed by introducing a new $SU(2)_R$ gauge symmetry,
with spatial parity symmetry 
is motivated to forbid the QCD $\bar\theta$ term at tree level.
  In particular, we focus on the minimal LR symmetric model, 
  which embeds the $SU(2)_L$ singlet right-handed quarks, $u_R$ and $d_R$, to the $SU(2)_R$ doublets, $Q_R \equiv (u_R, d_R)^T$. 
  Furthermore, 
  a $SU(2)_R$ doublet Higgs,  $H'$, and three flavors of the up-type and down-type vector-like quarks,  $U_L, U_R, D_L$ and $D_R$, have to be introduced.
  The matter contents are listed in
  Table~\ref{tab:matter contents}.
    \begin{table}[t]
        \centering
        \newcommand{\bhline}[1]{\noalign{\hrule height #1}}
        \renewcommand{\arraystretch}{1.5}
        \rowcolors{2}{gray!15}{white}
        \begin{tabular}{ c c c c c | c}
        \bhline{1 pt}
             &
            $SU(3)_C$      &
            $SU(2)_L$      &
            $SU(2)_R$      &
            $U(1)_{B-L}$   & 
            $U(1)_Y$\\
                  \hline    
            $Q_L^i\equiv(u_L^i,d_L^i)^T$      &
            $\bf 3$      &
            $\bf 2$      &
            $\bf 1$      &
            $1/6$ &
            $(1/6, 1/6)$
                  \\
            $Q_R^i\equiv(u_R^i,d_R^i)^T$      &
            $\bf 3$       &
            $\bf 1$      &
            $\bf 2$      &
            $1/6$ &
            $(2/3, -1/3 )$
                  \\
            $H$      &
            $\bf 1$      &
            $\bf 2$      &
            $\bf 1$      &
            $1/2$ &   
            $(1/2, 1/2)$
                  \\
            $H^\prime$      &
            $\bf 1$      &
            $\bf 1$      &
            $\bf 2$      &
            $1/2$ &
            $(1,0 )$
                  \\
            $U_L^a$      &
            $\bf 3$     &
            $\bf 1$      &
            $\bf 1$      &
            $2/3$ &  $2/3$
                  \\
            $U_R^a$      &
           $\bf 3$     &
            $\bf 1$      &
            $\bf 1$      &
            $2/3$ &  $2/3$
                  \\
            $D_L^a$      &
            $\bf 3$      &
            $\bf 1$      &
            $\bf 1$      &
            $-1/3$ & $-1/3$
                  \\
            $D_R^a$      &
            $\bf 3$      &
            $\bf 1$      &
            $\bf 1$      &
            $-1/3$ &  $-1/3$
                  \\
        \bhline{1 pt}
        \end{tabular}
        \caption{The matter contents and their gauge charges in the minimal LR symmetric model, where $U(1)_Y=T^R_3 \,+ \,U(1)_{B-L}$.
        The indices $i$ and $a$ represent the flavors for the doublet and singlet quarks, respectively.} 
        \label{tab:matter contents}
    \end{table}

    To solve the strong $CP$ problem, the spatial parity symmetry has to be extended to symmetrize
    the left-handed and right-handed sectors  as well as   the $SU(2)_L$ and $SU(2)_R$ gauge bosons, 
    \begin{eqnarray}
        \vec{x}&\longleftrightarrow&-\vec{x}\,, \nonumber\\
        W_\mu &\longleftrightarrow& W^{\prime \mu}\,, \\
        Q_L,~U_L,~D_L,~H &\longleftrightarrow& Q_R,~U_R,~D_R,~H^\prime\,,\nonumber
    \end{eqnarray}
    while the other gauge bosons are invariant. 
 The spontaneous violation 
 of the extended parity symmetry, 
 $SU(2)_R\times U(1)_{B-L} \to U(1)_Y$, 
 is caused by the vacuum expectation value (VEV) of $H^{\prime\,0}$,  $\langle H^\prime \rangle=(0,v^\prime)$.
 After the symmetry breaking, 
 the 
 $U(1)_Y$ gauge symmetry is generated with the gauge charge of 
 $U(1)_Y=T^R_3\, +\, U(1)_{B-L}$.
    The $SU(2)_R$ gauge bosons absorb the Nambu-Goldstone (NG) bosons in the doublet $H^\prime$  ($\varphi^{\prime\,+}$ and $\varphi^{\prime\,0}$)  to become massive states ($W^{\prime\,+}$ and $Z^{\prime}$). The physical neutral Higgs boson associated with this symmetry breaking is denoted as  $h^\prime$. 
    Then, the $SU(2)_L \times U(1)_Y$ gauge symmetry is broken to $U(1)_{\rm EM}$ by the VEV of $H^0$, $\langle H \rangle=(0,v)$. The $W^+$ and $Z$ bosons absorb the NG bosons in the doublet $H$ ($\varphi^{+}$ and $\varphi^{0}$), and the physical (SM) neutral Higgs boson  with this symmetry breaking is denoted as $h$.\footnote{ 
    The  $h$--$h'$ and $Z$--$Z'$ mixings are induced in the model at tree level, though the mixings are suppressed by $v/v^\prime$ and $(v/v^\prime)^2$, respectively \cite{Babu:1989rb}.
    Since we take $v/v^\prime\rightarrow 0$ in the calculation of the $\bar\theta$ parameter, they are ignored.}
    
    The resultant parity violation in nature comes from $v^\prime \neq v$. 
    Namely, we assume that soft parity breaking terms are contained 
    in the $H$ and $H^\prime$ Higgs potentials  which lead to $v'\gg v \neq 0$.  
    These two VEVs can be chosen as real and positive  without loss of generality. 
    Since the extended parity is a discrete symmetry, its spontaneous breaking leads to the formation of the domain walls, which dominate the energy density in the Universe. 
This domain wall problem can be naturally solved by the Planck suppressed higher-dimensional operators, which explicitly violate the parity symmetry \cite{Craig:2020bnv}.

 The Yukawa interactions
 and Dirac mass terms for the vector-like quarks are represented as
    \begin{equation}
        \begin{split}
            -{\cal L}_Y=&
            \overline{Q}_L^i x^{ia}_{u} U^a_{R}\tilde{H}
            + \overline{Q}_R^i x^{ia}_{u} U^a_{L}\tilde{H}^\prime
            +M^{a}_{u} \overline{U}_{L}^a U_R^a \\
            &+
            \overline{Q}_L^i x^{ia}_{d} D^a_{R}{H}
            + \overline{Q}_R^i x^{ia}_{d} D^a_{L}{H}^\prime
            +M^{a}_{d} \overline{D}_{L}^a D_R^a +{\rm h.c.} \,,            
         \end{split}
         \label{yukawa_and_dirac}
    \end{equation}
where $i=1$--$3$ is a flavor index for $SU(2)_{L/R}$ doublets, $a=1$--$3$ is that for the singlets (vector-like quarks), and
$\tilde{H}^{(\prime)}=\epsilon H^{(\prime)\star}$
($\epsilon_{12}=1$).
The LR symmetry requires
that the Yukawa $x^{i a}_{u/d}$ in the first two terms (in both lines) must be the same complex matrices, and the Dirac mass terms $M_u$ and $M_d$ must be  Hermitian. 
The Dirac mass terms $M_u^a$ and $M_d^a$ in Eq.~(\ref{yukawa_and_dirac}) are  diagonalized
to real and positive eigenvalues 
by the field redefinitions of the vector-like quarks.\footnote{%
One can also consider non-Hermitian vector-like quark mass matrices which correspond to soft parity breaking terms \cite{Babu:1989rb}. 
However, such contributions produce 
large quark EDM and radiative $\bar\theta$, and thus they are severely constrained from the EDM bounds \cite{Craig:2020bnv,deVries:2021pzl}. } 
The SM quark masses are realized by the seesaw mechanism such as higher dimensional operators induced by integrating out the vector-like quarks.\footnote{%
One can also extend the lepton sector that is insensitive to the QCD $\bar\theta$ term.
If one considers $SU(5)_L \times SU(5)_R$ grand unification \cite{Davidson:1987mi,Davidson:1987mh,Lee:2016wiy}, vector-like neutral leptons are absent and 
the neutrinos keep massless at the tree level.
Interestingly, 
suitable Dirac neutrino masses are generated from the two-loop radiative corrections \cite{Babu:1988yq} with predicting a nonzero $\Delta N_{\rm eff}$ \cite{Babu:2022ikf}.
Furthermore, an $\mathcal{O}(10)$\,keV sterile neutrino dark matter with the leptogenesis mechanism can be incorporated \cite{Dror:2020jzy,Dunsky:2020dhn}.}

Before discussing the mass matrices in detail in the next section, 
let us count on the number of physical $CP$ phases in this model.
The Yukawa couplings $x_{u/d}^{ia}$ are $3 \times 3$ complex matrices.  
Nine real parameters are removed from the Yukawa matrices by field redefinition of $Q_{L/R}^i$ as $Q_{L/R}^i\rightarrow U^{ij} Q_{L/R}^j$ with  a unitary matrix $U$.
Furthermore, phase redefinition of $U_{L/R}^a$ and $D_{L/R}^a$ removes five phases in the total. 
A remaining phase rotation corresponds to the baryon number conservation, and it does not change $x_u$ and $x_d$. Thus, $x_u$ and $x_d$ have a total of 22 physical real parameters.
We parametrize these 22 parameters as 
   \begin{align}
   \begin{aligned}
      x_u&=
      \overline{\Phi}^\dagger(\theta_{d3},\theta_{d8})\,
       V_Q\, \overline{\Phi}(\theta_{u3},\theta_{u8})\,
    \bar{x}_u V_U\,,\\
         x_d&= \bar{x}_d  V_D \,,
         \label{parameterization1}
    \end{aligned}     
    \end{align}
with
      \begin{eqnarray}
      \overline{\Phi}(\theta_{3},\theta_{8})&\equiv&
      \exp(i\tau_3 \theta_3)\exp(i\tau_8 \theta_8)\,,
    \end{eqnarray}
    where $\tau_3$ and $\tau_8$ are the third and eighth Gell-Mann matrices.
Here,  $\bar{x}_{u/d}$ are real diagonal matrices and $V_Q$, $V_U$, and $V_D$ are CKM-like unitary matrices
which have three rotation angles and one $CP$-violating phase. %
 It is found that there are seven $CP$-violating phases in this model ($\theta_{u3/u8}$, $\theta_{d3/d8}$, and three phases in $V_{Q/U/D}$).
When the Dirac masses are assumed to be universal such as $M_u^a=M_u$ and $M_d^a=M_d$ for $a=1$--$3$, the parameters  $\theta_{u3/u8}$,
    $\theta_{d3/d8}$,  $V_{U/D}$ become unphysical and only $V_Q$ remains physical.   

In this paper, we assume that $v^\prime \lsim M_{q}^{a}$ and will utilize expansions by $v'/M_{q}^{a}$. 
This inequality is motivated because the seesaw mechanism may explain the SM fermion mass hierarchy naturally.
On the other hand,
if $v^\prime \gg M_{q}^{a}$, a copy of the SM fermions has a mass spectrum similar to the SM fermions, which 
spread over five orders of magnitude. A new naturalness problem might appear in such a model, 
but the QCD $\bar\theta$ term  is suppressed by $M_{q}^{a}/v^\prime$ since only the CKM phase survives in a limit of $M_{q}^{a}\rightarrow 0$. In the following, we will consider the case of $v^\prime \lsim M_{q}^{a}$.

\subsection{Parametrization of Yukawa coupling constants}

In this section, we show the quark mass matrices and define the mass eigenstates. In the mass matrices   the Yukawa coupling constants, $x_u$ and $x_d$, appear, 
and we have to determine them from the observed quark masses and CKM matrix in order to evaluate the radiative $\bar\theta$ parameter.
We give the 
parameterization of the Yukawa coupling constants assuming
the SM quark masses are given by the seesaw mechanism with  $v^\prime \lesssim M_{q}^{a}$.

From Eq.~(\ref{yukawa_and_dirac}), the quark mass matrices in the flavor eigenstates are given as 
    \begin{align}
         -{\cal L}_M & =
  \left(\overline{u}_L^i, \overline{U}_L^a\right)
            \begin{pmatrix}
                0&x^{ib}_{u} v\\
                x^{\dagger aj}_{u} v^{\prime} & M^{a}_{u} \delta^{ab}
            \end{pmatrix}
            \begin{pmatrix}
                u_R^j \\
                U_R^b
            \end{pmatrix} + \left(\overline{d}_L^i, \overline{D}_L^a\right)
            \begin{pmatrix}
                0&x^{ib}_{d} v\\
                x^{\dagger aj}_{d} v^{\prime} & M^{a}_{d} \delta^{ab}
            \end{pmatrix}
            \begin{pmatrix}
                d_R^j \\
                D_R^b
            \end{pmatrix} 
            +{\rm h.c.} \nonumber \\
&\equiv \overline{\mathcal{U}}_L^p \mathcal{M}_u^{(0) pq} \mathcal{U}_R^q + \overline{\mathcal{D}}_L^p \mathcal{M}_d^{(0) pq} \mathcal{D}_R^q+{\rm h.c.} \,,  
    \end{align}
    where   ${\mathcal{U}}_{L/R}^p $ and ${\mathcal{D}}_{L/R}^p $ ($p,q=1,\cdots,6$) are the up- and down-type flavor eigenstates, 
    and $v\simeq 174.1\,$GeV.
    Here, 
    $M^{a}_{u/d}$ are real diagonal, while $x^{ia}_{u/d}$ are complex matrices.
    It is obvious that $\text{arg}\, \text{det} \,{\cal M}^{(0)}_{u/d}=0$.
    Then, 
    the $6 \times 6$ fermion mass matrices are  diagonalized by  bi-unitary matrices, $V_{qL}$ and $V_{qR}$, as
    \begin{eqnarray}
      {\cal M}_q^{(0) pq}&=&V_{qL}^{\dagger pP} \bar{M}_q^{P} V_{qR}^{Pq} \quad \text{for}~q=u \text{~and~}d\,,
      \label{diagnalizing}
    \end{eqnarray}
    with diagonal mass matrices  $\bar{M}_q$.  The mass eigenstates, ${\mathcal{U}}_{ML/R}^P $ and ${\mathcal{D}}_{ML/R}^P $  for $P=1$--$6$, 
    are given as
\begin{align}
      \begin{aligned}
      {\cal U}_{ML}^{P}&=V_{uL}^{Pp} {\cal U}_{L}^{p}\,,
      \qquad 
      {\cal U}_{MR}^{P}=V_{uR}^{Pp} {\cal U}_{R}^{p}\,,\\
      {\cal D}_{ML}^{P}&=V_{dL}^{Pp} {\cal D}_{L}^{p}\,,
      \qquad 
      {\cal D}_{MR}^{P}=V_{dR}^{Pp} {\cal D}_{R}^{p}\,.
      \end{aligned}
      \label{eq:diagonalization}
 \end{align}

    It is difficult to reconstruct the model parameters from the experimental data in general. Here we assume that $v^\prime \lsim M_{q}^{a}$ and we take leading terms in the expansion of $v^\prime/M^{a}_{q}$ for the quark mass eigenvalues. 
    In this expansion, the SM quark masses are given
    by the following seesaw relation
    \begin{equation}
        V_q^{\dagger iI} \frac{m_q^{I}}{v} V_q^{Ij}= x_{q}^{ia}\frac{v^\prime}{M_{q}^{a}}x_{q}^{\dagger aj}\,,
        \label{eq:seesaw}
    \end{equation}
with a $3 \times 3$ unitary matrix $V_{q} $ and  
$I=1$--$3$, while 
the heavy quark masses 
are given by $M_q^a$, for $q=u$ and $d$.

Now let us rewrite Eq.~(\ref{eq:seesaw}) as
    \begin{eqnarray}
      {\boldsymbol I}_3&=&\left(\frac{\sqrt{v}}{\sqrt{m_q}}V_q x_{q}\frac{\sqrt{v^\prime}}{\sqrt{M_{q}}}\right)
     \left(\frac{\sqrt{v^\prime}}{\sqrt{M_{q}}}x^{\dagger}_{q} V^\dagger_q\frac{\sqrt{v}}{\sqrt{m_q}}\right)
     \equiv U_q U^\dag_q\,,
    \end{eqnarray}
    where 
  ${\boldsymbol I}_3$ is a unit matrix in the three-dimensional space and   
    we ignore the indices of matrices.\footnote{
A similar parameterization technique, referred to as the  Casas-Ibarra parameterization, 
is applied in the minimal seesaw model~\cite{Casas:2001sr, Ellis:2002fe}.
    } 
Thus, the Yukawa matrices $x_q$ are given with a unitary matrix $U_q$ by
    \begin{eqnarray}
      x_{q}=V^{\dagger }_q \frac{\sqrt{m_q}}{\sqrt{v}} U_q\frac{\sqrt{M_{q}}}{\sqrt{v^\prime}}\,.
     \end{eqnarray}
According to the previous section,
one can remove some unphysical parameters in $U_q$ and $V_q$ by the field redefinitions.
Then, we get
    \begin{align}
        \begin{aligned}
            x_{u}& =V_{\rm CKM}^\dagger\frac{\sqrt{m_u}}{\sqrt{v}}\overline{\Phi}(\theta_{u3}, \theta_{u8})V_U \frac{\sqrt{M_{u}}}{\sqrt{v^\prime}}\,, \\
            x_{d}& =\frac{\sqrt{m_d}}{\sqrt{v}}\overline{\Phi}(\theta_{d3}, \theta_{d8})V_D \frac{\sqrt{M_{d}}}{\sqrt{v^\prime}}\,,
        \end{aligned}
        \label{parametrization2}
    \end{align}
    where $V_{\rm CKM}(\equiv V_u V_d^\dagger)$ corresponds to the CKM matrix in the SM. 
Here, $V_{U/D}$ are CKM-like unitary matrices
with three mixing angles and one $CP$-violating phase, though they are different from those in Eq.~(\ref{parameterization1}). 
Now we have seven physical $CP$-violating phases ($\theta_{u3/u8}$, $\theta_{d3/d8}$, and three phases in $V_{U/D}$ and $V_{\rm CKM}$), which is consistent with our previous counting, and all phases can be $\mathcal{O}(1)$ under the extended parity symmetry.

Since we assume that $v^\prime \lsim M_{q}^{a}$, the $6 \times 6$ diagonalization matrices in Eq.~\eqref{eq:diagonalization} are given  of leading terms in the expansion of $v^\prime/M^{a}_{q}$ as
\begin{align}
    \begin{aligned}
      V_{uL}&\simeq 
      \begin{pmatrix}
        -V_{\rm CKM} &V_{\rm CKM} x_u \frac{v}{M_u}\\
        \frac{v}{M_u}x_u^{\dagger} &{\boldsymbol I}_3
      \end{pmatrix}\,,\\
      V_{uR}&\simeq 
      \begin{pmatrix}
        V_{\rm CKM} &-V_{\rm CKM} x_u \frac{v^\prime}{M_u}\\
        \frac{v^\prime }{M_u}x_u^{\dagger} &{\boldsymbol I}_3
      \end{pmatrix}\,,\\
      V_{dL}&\simeq 
      \begin{pmatrix}
        -{\boldsymbol I}_3 & x_d \frac{v}{M_d}\\
        \frac{v}{M_d}x_d^{\dagger} &{\boldsymbol I}_3
      \end{pmatrix}\,,\\
      V_{dR}&\simeq 
      \begin{pmatrix}
        {\boldsymbol I}_3 &- x_d \frac{v^\prime}{M_d}\\
        \frac{v^\prime }{M_d}x_d^{\dagger} &{\boldsymbol I}_3
      \end{pmatrix}\,,
    \end{aligned}
 \label{mixing_matrix}
\end{align}
where $x_q$ are given by Eq.~(\ref{parametrization2}). 
Here, the diagonal eigenvalue matrices $\bar{M}_q$  are given as,
\begin{align}
\bar{M}_q^P =\text{diag}(m_q^I, M_q^a)
\quad \text{for}~q=u \text{~and~}d\,.
\label{eq:massP}
\end{align}

\subsection{Quark EDMs}

Before discussing the radiative $\bar\theta$ parameter in the minimal LR symmetric model,
let us comment on contributions to quark (chromo) EDMs. As long as the vector-like  mass matrices are Hermitian, 
the quark EDMs vanish completely at one-loop level. The $W^{(\prime)\pm}$ contribution at one-loop level vanishes trivially since the chirality is conserved in the diagrams. 
On the other hand, the one-loop quark EDM contributions from neutral Higgses and $Z^{(\prime)}$ may have a chirality flip in the diagrams.
Nevertheless, they also vanish because the extended parity symmetry restricts the product of two vertices to be strictly real, as shown in  Ref.~\cite{Craig:2020bnv}.

\section{Confirmation of vanishing QCD  \texorpdfstring{\boldmath{$\theta$}}{theta} parameter in  two-loop order}
\label{sec:proof}

The parity symmetry is spontaneously broken by the $\langle H^\prime \rangle~(\gg \langle H\rangle)$ in the LR symmetric model.
Since $\text{arg}\, \text{det} \,{\cal M}^{(0)}_{u/d}=0$ holds, fermion one-loop contributions to the QCD $\bar\theta$ term remain zero.
However, it is expected that 
fermion-loop diagrams at a higher than one-loop level
would generate it.
In this section, we show fermion two-loop contributions to the QCD $\bar\theta$ term still vanish. 

Integrating out quarks, the following higher-dimensional operators are expected to be generated,
\begin{eqnarray}
{\cal L}_{\rm eff}&=& 
\sum_{n=1} 
C_n \frac{|H^\prime|^{2n}-|H|^{2n}}{M_q^{2n}}
\frac{\alpha_s}{8\pi}
G^{a}_{\mu\nu}\tilde{G}^{a\mu\nu}\,,
\label{HDO}
\end{eqnarray}
with $M_q$ as the scale of vector-like quark masses, 
and the QCD $\bar\theta$ term are induced by the spontaneous parity symmetry breaking $\langle H^\prime \rangle~(\gg \langle H\rangle)$,
\begin{eqnarray}
    \bar{\theta}&=&
  \sum_{n=1} C_n \left(\frac{\langle H^\prime \rangle}{M_q}\right)^{2n}.
\end{eqnarray}
We evaluate the Wilson coefficients of the operators $C_n$ in the following. 

First, we consider the contributions to the QCD $\bar\theta$ term 
at two-loop level coming from an exchange of the $W^{\prime\,\pm}$ boson. 
The two-loop fermion bubble diagrams mediated by the $W^{\prime\,\pm}$ boson under the gluon field-strength background conserve chirality in the fermion line, and then it is proportional to
\begin{eqnarray}
    V_{dR}^{Pi}V_{uR}^{\dagger iQ}
    V_{uR}^{Qj}V_{dR}^{\dagger jP} f\left[(\bar{M}_d^P)^2,(\bar{M}_u^Q)^2, m_{W^{\prime}}^2\right]\,,
    \label{eq:Wp_loop}
\end{eqnarray}
where $i$ and $j$ run $1$--$3$ as the flavor index for the $SU(2)_R$ doublet, while
$P$ and $Q$ run $1$--$6$ 
for the quark mass eigenstates.
Here,
a two-loop function $f[(\bar{M}_d^P)^2,(\bar{M}_u^Q)^2,m_{W^{\prime}}^2]$ 
is a real function. Since $ \left(V_{dR}^{Pi}V_{uR}^{\dagger iQ}
    V_{uR}^{Qj}V_{dR}^{\dagger jP} \right)^\ast =  V_{dR}^{Pj}V_{uR}^{\dagger jQ}
    V_{uR}^{Qi}V_{dR}^{\dagger iP}$
    and it corresponds to Eq.~\eqref{eq:Wp_loop} by an exchange of $i \leftrightarrow j$, the contribution is real so that it does not generate the QCD $\bar\theta$ term.

The reason why the exchange of the $W^{\prime\,\pm}$ boson does not contribute to  the QCD $\bar\theta$ term at two-loop level is clear. However, the above discussion is based on the structure of the mixing matrices in the contribution, not on the structure of Lagrangian parameters, such as $x_{u/d}$ and $M_{u/d}$, and then, it is unclear what is required to generate the QCD $\bar\theta$ term in higher-order diagrams. We make it clear by explicit calculation of the loop diagrams in the following.

In the unitary gauge, the lowest dimension operator ($n=1$) in Eq.~(\ref{HDO}) might come from diagrams which include the longitudinal mode of the $W^{\prime\,\pm}$ boson.
It is because the propagator is proportional to $k^\mu k^\nu/m_{W^\prime}^2$ ($k^\nu$ the momentum of  the $W^{\prime\,\pm}$ boson),
and it could give the lowest order contribution with regard to $v^{\prime \,2}$. The Yukawa coupling constants $x_u$ and $x_d$ are multiplied with the Higgs VEVs in the mixing matrices as in Eq.~(\ref{mixing_matrix}). 

In our calculation, we adopt the $R_\xi$ gauge with the  Feynman-'t~Hooft gauge $\xi=1$ (for $SU(2)_L \times SU(2)_R \times U(1)_{B-L}$ gauge), in order to avoid the messy calculation in the unitary gauge. The lowest dimension operator ($n=1$) in Eq.~(\ref{HDO}) could arise the charged NG boson exchange  $\varphi^{\prime\,\pm}$ in this gauge. The charged NG boson is absorbed by $W^{\prime\,\pm}$ boson in the Higgs mechanism, and its mass, $m_{\varphi^\prime}$, is equal to the $W^{\prime\,\pm}$ mass. The charged NG boson interactions are given as
    \begin{eqnarray}
      -{\cal L}_{\varphi^{\prime \pm}} &=&
\overline{u}_R^i x_d^{ia} D^a_{L}\varphi^{\prime +}
            - \overline{d}_R^i x_u^{ia} U^a_{L} \varphi^{\prime\,-}
      +{\rm h.c.}\nonumber\\
    &=&
    (x_d^{ia} V_{uR}^{Pi} V_{dL}^{\star Qa}) \, \overline{\cal U}_{MR}^{P} {\cal D}^{Q}_{ML} \varphi^{\prime +}
    -  (x_{u}^{\star ia } V_{uL}^{Pa} V_{dR}^{\star Qi} ) \, \overline{\cal U}_{ML}^{P} {\cal D}^{Q}_{MR} \varphi^{\prime +}
    +{\rm h.c.}  \,.
    \end{eqnarray}
Both the left- and right-handed quarks are coupled with the charged NG boson.  We will show that the charged NG boson diagrams at two-loop level do not contribute to the QCD $\bar\theta$ term. 

    \begin{figure}[t]
        \centering
        \begin{subfigure}{1.0 \textwidth}
        \includegraphics[width=1.0\linewidth]{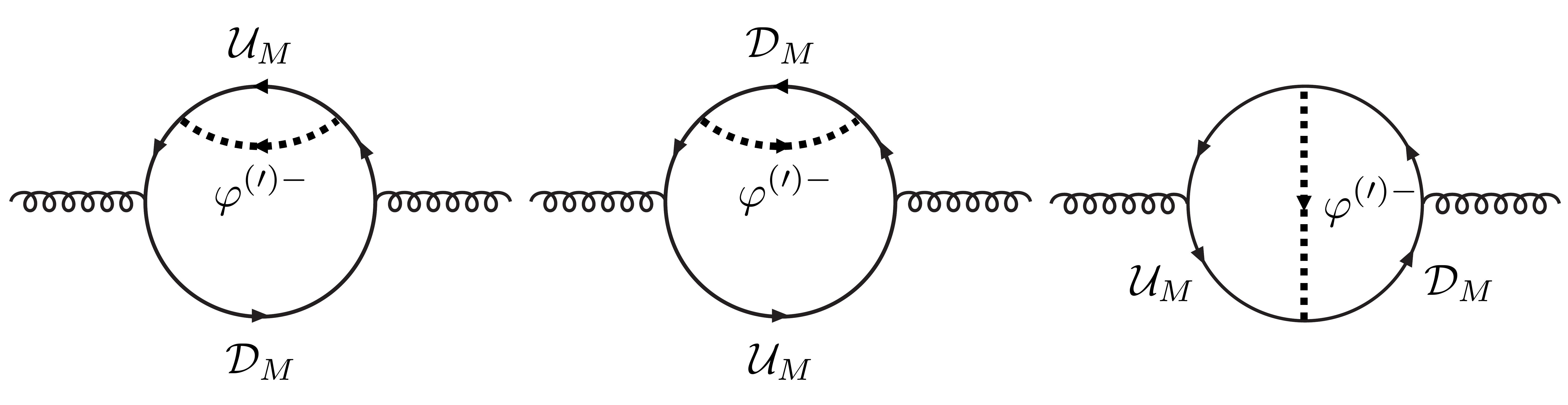}
        \caption{diagram $A$  \qquad \qquad \qquad \qquad (b) diagram $B$ \qquad \qquad \qquad \qquad (c) diagram $C$}
        \end{subfigure}
        \caption{The charged NG boson contributions to the QCD $\bar\theta$ term at two-loop level.
        }
        \label{fig:charged boson exchange}
    \end{figure}
Three diagrams (dubbed as diagrams $A$, $B$ and $C$) in Fig.~\ref{fig:charged boson exchange} could give contributions to the $\bar\theta$ parameter.
By using the Fock-Schwinger gauge method (for $SU(3)_C$ gauge) in Sec.~\ref{sec:FSG}, we obtain 
\begin{align}
      \delta \theta|_{A}&=
      \frac{1}{8\pi^2}{\rm Im}\left(x_u^{ia} V_{uL}^{\star Pa} V_{dR}^{Qi} x_d^{jb} V_{uR}^{Pj} V_{dL}^{\star Qb}\right)\bar{M}_{u}^P\bar{M}_{d}^Q \, \bar{I}_{(1;3)}\left[(\bar{M}_{u}^P)^2;(\bar{M}_{d}^Q)^2;m_{\varphi^\prime}^2\right]\,,\label{delta_thetaA}\\
      \delta \theta|_{B}&=
      \frac{1}{8\pi^2}{\rm Im}\left(x_u^{ia} V_{uL}^{\star Pa} V_{dR}^{Qi} x_d^{jb} V_{uR}^{Pj} V_{dL}^{\star Qb}\right)\bar{M}_{u}^P\bar{M}_{d}^Q \, \bar{I}_{(3;1)}\left[(\bar{M}_{u}^P)^2;(\bar{M}_{d}^Q)^2;m_{\varphi^\prime}^2\right]\,,\label{delta_thetaB}\\
      \delta \theta|_{C}&=
      \frac{1}{8\pi^2}{\rm Im}
      \left(x_u^{ia} V_{uL}^{\star Pa} V_{dR}^{Qi} x_d^{jb} V_{uR}^{Pj} V_{dL}^{\star Qb}\right)
      \bar{M}_{u}^P\bar{M}_{d}^Q \,I_{(2;2)}\left[(\bar{M}_{u}^P)^2;(\bar{M}_{d}^Q)^2;m_{\varphi^\prime}^2\right]\,,
      \label{delta_thetaC}
\end{align}
where the two-loop functions $\bar{I}_{(1;3)}$, $\bar{I}_{(3;1)}$, and $I_{(2;2)}$ are defined in Appendix~\ref{app:loop}. 
In the above evaluation, we pick up contributions proportional to both $x_u$ and $x_d$, which are also proportional to the quark masses in the mass eigenstate propagators.
The terms proportional to $x_u$ and $x_u^\star$ (or $x_d$ and $x_d^\star$) is real. 

Here, we use the dimensional regularization $(d=4-2\epsilon)$ for loop momentum integrals
and the partial diagrams produce UV  divergence.
However, 
the contributions from diagrams $A$ and $B$, proportional to $\bar{I}_{\epsilon{(1;3)}}$ and $\bar{I}_{\epsilon{(3;1)}}$ in Eq.~\eqref{eq:UVdiv}, respectively, vanish, so that
the correction to the $\bar\theta$ parameter is finite and scale-independent. 
 The UV divergent parts ($1/\epsilon$ terms) in  $\bar{I}_{\epsilon{(1;3)}}$ and $\bar{I}_{\epsilon{(3;1)}}$ cancel out since
    \begin{eqnarray}
      {\rm Im}\left(x_u^{ia} V_{uL}^{\star Pa} V_{dR}^{Qi} x_d^{jb} V_{uR}^{Pj} V_{dL}^{\star Qb}\right)\frac{\bar{M}_{u}^P}{\bar{M}_{d}^Q}
      &=&{\rm Im}\left(x_u^{\star i a}x_{u} ^{i a}\right)=0\,,\\
      {\rm Im}\left(x_u^{ia} V_{uL}^{\star Pa} V_{dR}^{Qi} x_d^{jb} V_{uR}^{Pj} V_{dL}^{\star Qb}\right)\frac{\bar{M}_{d}^Q}{\bar{M}_{u}^P}
      &=&{\rm Im}\left(x_d^{\star ia } x_{d} ^{i a}\right)=0\,.
    \end{eqnarray}  
To derive the above equations we use the following equations,
    \begin{eqnarray}
      [({\cal M}_q^{(0)})^{-1}]^{pq}&=&
      V_{qR}^{\dagger pP}(\bar{M}_q^{-1})^{P} V_{qL}^{Pq}
     \nonumber \\
  &= &
    \begin{pmatrix}
-\frac{1}{vv^\prime}(x_q^{\dagger})^{-1}_{ic}M^c_q(x_{q})^{-1}_{cj}      
      &\frac1{v^\prime}(x_q^{\dagger})^{-1}_{ib}\\
      \frac1v(x_{q})^{-1}_{aj} &0
    \end{pmatrix}
    \quad \text{for}~q=u \text{~and~}d\,,
      \label{inverse_matrix}
    \end{eqnarray}
 in addition to Eq.~(\ref{diagnalizing}) with $(\bar{M}_q^P)^\ast = \bar{M}_q^P$.
 Furthermore, we observed that the following combinations also vanish\footnote{%
 The factors $1/M \log M$ ($M$:~quark mass) in Eqs.~(\ref{mlogm1}) and (\ref{mlogm2}) correspond to the $O(\epsilon)$ term in Eq.~(\ref{FGgauge_oneloop}) when changing $d^4p$ to $d^dp$ $(d=4-2\epsilon)$. They become  $O(\epsilon^0)$ since $1/\epsilon$ comes from the quark self-energy subdiagrams in diagrams $A$ and $B$. Equations~(\ref{mlogm1}) and (\ref{mlogm2}) can be perturbatively proved by assuming the off-diagonal terms in the quark mass matrices are small. The similar trick is also used around  Eq.~\eqref{eq:LOvanish}.
 We also checked Eqs.~(\ref{mlogm1}) and (\ref{mlogm2}) numerically.}
     \begin{eqnarray}
      {\rm Im}\left(x_u^{ia} V_{uL}^{\star Pa} V_{dR}^{Qi} x_d^{jb} V_{uR}^{Pj} V_{dL}^{\star Qb}\right)\frac{\bar{M}_{u}^P}{\bar{M}_{d}^Q} \log \bar{M}_{d}^Q 
      &=&0\,,
       \label{mlogm1}\\
      {\rm Im}\left(x_u^{ia} V_{uL}^{\star Pa} V_{dR}^{Qi} x_d^{jb} V_{uR}^{Pj} V_{dL}^{\star Qb}\right)\frac{\bar{M}_{d}^Q}{\bar{M}_{u}^P} \log \bar{M}_{u}^P
      &=&0\,.
      \label{mlogm2}
    \end{eqnarray}  
Therefore, the second terms of  $\bar{I}_{\epsilon{(1;3)}}$ and $\bar{I}_{\epsilon{(3;1)}}$ in Eq.~\eqref{eq:UVdiv} also do not affect the $\bar\theta$ parameter.  
On the other hand, the loop function $I_{(2;2)}$ in  Eq.~(\ref{delta_thetaC}) is UV finite. 

Similar to the $\varphi^{\prime\,\pm}$ contribution,     
    the contribution from the SM charged NG boson $\varphi^\pm$, absorbed into $W^\pm$, is derived by replacing $L(R)$ with $R(L)$, and $m_{\varphi^\prime}^2$ with $m_{\varphi}^2$ in the above formulae. 
Furthermore,  $(-1)$ is multiplied since
the chiralities of circulating fermions are opposite to diagrams of Fig.~\ref{fig:charged boson exchange}. 
    It means that, if one sets $v=v^\prime$ corresponding to the LR symmetric limit, those two contributions of $\varphi^{\prime\,\pm}$ and $\varphi^{\pm}$ cancel each other.  

The diagrams $A$ and $B$ correspond to the one-loop correction to the fermion mass terms.  The two-loop function $\bar{I}_{(3;1)}(x_1;x_2;x_3)$ is expressed as  
  \begin{eqnarray}
      \bar{I}_{(3;1)}(x_1;x_2;x_3)&=&(16\pi^2\mu^{2\epsilon})\int\frac{d^d p}{i(2\pi)^d}
        \frac{1}{(p^2-x_1)^3} F_0(p^2,x_2,x_3)\,,
    \end{eqnarray}
  where $F_0(p^2,x_2,x_3)$ is a loop function of one-loop diagrams for the fermion mass correction,
    \begin{eqnarray}
      F_0(p^2,x_2,x_3)&=&\int^1_0dz \log\frac{-z(1-z)p^2+z x_2+(1-z) x_3}{Q^2}\,,
    \end{eqnarray}
with $Q^2\equiv4\pi \mu^2 {\rm e}^{-\gamma_E}$ and $\mu$ is the renormalization scale.
($\bar{I}_{(1;3)}(x_1;x_2;x_3)$ also has a similar expression, see Appendix~\ref{app:loop}.) 
$\bar{I}_{(3;1)}(x_1;x_2;x_3)$ has an IR-singular behavior when $x_1\ll x_3$ as 
    \begin{eqnarray}
      \bar{I}_{(3;1)}(x_1;x_2;x_3)&\simeq &-\frac1{2x_1}F_0(0,x_2,x_3) \,,
    \end{eqnarray}
while small $x_2$ and $x_3$ do not lead to IR singularities.
This behavior is expected.
It is because if a fermion with real mass $m_f$ gets a constant radiative correction to the fermion mass $m_f + \delta m_f$, the correction to the $\bar\theta$ parameter
is given by $\delta \theta\simeq -{\rm Im}(\delta m_f)/m_f$, see Eq.~(\ref{FGgauge_oneloop}).\footnote{
In Eq.~(\ref{FGgauge_oneloop}), the loop function and the chirality flip lead to $\sim m_f/m_f^2=1/m_f$, and then $\delta \theta\simeq -{\rm Im}(\delta m_f)/m_f$.
}
However, this evaluation of $\delta \theta$ is justified only when the correction to the fermion mass term is independent of fermion momentum.

The IR-singular behaviors of the SM fermion masses in $\bar{I}_{(3;1)}$ and $\bar{I}_{(1;3)}$ are not physical in $\delta \theta|_A$ and $\delta \theta|_B$ 
in Eqs.~(\ref{delta_thetaA}) and (\ref{delta_thetaB}),  
and they can be removed indeed using Eq.~(\ref{inverse_matrix}) as
    \begin{align}
        \begin{aligned}
                \delta \theta|_{A}\simeq &\,\frac{1}{8\pi^2}{\rm Im} \left[x_u^{ia}x_d^{jb} \left( V_{uL}^{\star A a} \bar{M}_{u}^{A}V_{uR}^{Aj} \right) \left( V_{dR}^{Bi} \frac1{\bar{M}_{d}^{B}} V^{\star Bb}_{dL} \right) \right] \\
                &\times \left[ (\bar{M}_{d}^{B})^2 \bar{I}_{(1;3)} \left((\bar{M}_{u}^{A})^2; (\bar{M}_{d}^{B})^2; m^2_{\varphi^\prime}\right) -
                (\bar{M}_{d}^{B})^2
                \bar{I}_{(1;3)}\left(0;(\bar{M}_{d}^{B})^2;m^2_{\varphi^\prime}\right)\right. \\
                &\qquad \left. +\frac12 F_0 \left(0,(\bar{M}_{u}^{A})^2,m^2_{\varphi^\prime}\right) - \frac12 F_0 \left(0,0,m^2_{\varphi^\prime}\right) \right] \\
                &+ \frac{1}{8\pi^2}{\rm Im} \left[x_u^{ia}x_{u}^{\star ja}x_d^{jb}v^\prime \left(V_{dR}^{Ai} \frac1{\bar{M}_d^{A}} V^{\star Ab}_{dL} \right) \right] \\
                &\times \left[ (\bar{M}_d^{A})^2 \bar{I}_{(1;3)} \left(0;(\bar{M}_d^{A})^2 ; m^2_{\varphi^\prime} \right) + \frac12 F_0 \left( 0 , 0 , m^2_{\varphi^\prime}\right) \right]\,,
                \label{two-loop-exact1A}
            \end{aligned} \\
            \begin{aligned}
                \delta \theta|_{B} \simeq &\, \frac{1}{8\pi^2} {\rm Im} \left[ x_u^{ia} x_d^{jb} \left( V_{uR}^{Aj} \frac1{\bar{M}_u^{A}}V^{\star Aa}_{uL} \right) \left( V^{\star Bb}_{dL} \bar{M}_d^{B} V_{dR}^{Bi} \right) \right] \\
            &\times \left[ ( \bar{M}_u^{A})^2 \bar{I}_{(3;1)} \left( (\bar{M}_u^{A})^2 ; (\bar{M}_d^{B})^2 ; m^2_{\varphi^\prime} \right) - 
            (\bar{M}_{u}^{A})^2
            \bar{I}_{(3;1)} \left( (\bar{M}_u^{A})^2 ; 0 ; m^2_{\varphi^\prime} \right) \right. \\
            &\qquad \left. + \frac12 F_0 \left( 0 , (\bar{M}_d^{B})^2 , m^2_{\varphi^\prime} \right) - \frac12 F_0 \left( 0 , 0 , m^2_{\varphi^\prime} \right) \right] \\
            &+ \frac{1}{8\pi^2}{\rm Im} \left[x_u^{ia} x_d^{jb} x_{d}^{\star ib} v^\prime \left(V_{uR}^{Aj} \frac1{\bar{M}_u^{A}} V_{uL}^{^\star Aa}\right)\right] \\
            &\times \left[ (\bar{M}_u^{A})^2 \bar{I}_{(3;1)} \left(
            (\bar{M}_u^{A})^2 ; 0 ; m^2_{\varphi^\prime} \right) + \frac12 F_0 \left( 0 , 0 , m^2_{\varphi^\prime} \right) \right] \,,
            \label{two-loop-exact1B}
        \end{aligned}
    \end{align}
where $A$ and $B$ run $4$--$6$ 
as the heavy quark mass eigenstates, see Eq.~\eqref{eq:massP}.  Here, the SM quark masses in the loop function are taken to be zero.
    
    On the other hand, 
    the contribution of diagram $C$ is not associated with the correction to the quark masses, and then it is a new type contribution to the $\bar\theta$ parameter. It is suppressed by the heavier fermion or $\varphi^{\prime\,\pm}$ masses. 
    By taking the SM quark masses to be zero in the loop function (see Appendix~\ref{app:loop}), it is given as 
    \begin{align}
        \delta\theta|_C \simeq  \frac{1}{8\pi^2} {\rm Im} \left[ x_u^{ia} x_d^{jb} \left( V_{uL}^{\star Aa} \bar{M}_u^{A} V_{uR}^{Aj} \right) \left( V_{dL}^{\star Bb} \bar{M}_d^{B} V_{dR}^{Bi} \right) \right] {I}_{(2;2)} \left( (\bar{M}_u^{A})^2 ; (\bar{M}_d^{B})^2 ; m^2_{\varphi^\prime} \right)\,,
        \label{two-loop-exact2}
    \end{align}
where $A$ and $B$ run $4$--$6$ 
as the heavy quark mass eigenstates. It is found that
the diagram $C$ does not contain any IR-singular behavior, unlike the diagrams $A$ and $B$.

    When $v^\prime \lsim M_q^a$,
    the leading contributions of $\mathcal{O}(v^{\prime 2}/(M_q^a)^2)$ 
    are given 
     as 
    \begin{align}
      \delta\theta|_A&\simeq \frac{1}{8\pi^2}
     {\rm Im}
     \left[(A_u^a)^{ij} (A_d^{b})^{ji}\right]
        \left\{v^{\prime2}\,\bar{I}_{(1;3)}
        \left[ (M_{u}^a)^2;(M_{d}^b)^2;m^2_{\varphi^\prime}\right]
        + \frac{v^{\prime2}}{2(M_{d}^b)^2} F_0\left[0,(M_{u}^a)^2,m^2_{\varphi^\prime}\right]\right\}\,,\\
        \delta \theta|_{B}&\simeq 
      \frac{1}{8\pi^2}{\rm Im}
        \left[(A_u^a)^{ij} (A_d^{b})^{ji}\right]
        \left\{v^{\prime2}\,\bar{I}_{(3;1)}
        \left[ (M_{u}^a)^2;(M_{d}^b)^2;m^2_{\varphi^\prime}\right]
        + \frac{v^{\prime2}}{2 (M_{u}^a)^2}F_0\left[0,(M_{d}^b)^2,m^2_{\varphi^\prime}\right]\right\}\,,\\
      \delta\theta|_C&\simeq \frac{1}{8\pi^2}
    {\rm Im}
        \left[(A_u^a)^{ij} (A_d^{b})^{ji}\right] v^{\prime2}\,
        {I}_{(2;2)} \left[(M_{u}^a)^2;(M_{d}^b)^2;m^2_{\varphi^\prime}\right]\,,
    \end{align}
with a Hermitian matrix $A_q^a,$ 
    \begin{align}
    (A_q^a)^{ij}\equiv x_q^{ia}x_q^{\star ja}
    \quad \text{for}~q=u, d \text{ and~not~sum~}a\text{~index}\,.
    \end{align}    
This can be derived from the above formulae by
    \begin{eqnarray}
      (V^\dagger_{qL})^{aA}
      \frac{\bar{M}_{q}^{A}}{p^2-(\bar{M}_{q}^{A})^2}
      (V_{qR})^{Ai}
      &\rightarrow&
      \frac{x_q^{\dagger ai}v^\prime}{p^2-(M_{q}^a)^2}\,,\\
      (V^\dagger_{qL})^{aA}
      \frac{\bar{M}_{q}^{A}}{[p^2-(\bar{M}_{q}^{A})^2]^2}
      (V_{qR})^{Ai}
      &\rightarrow&
      \frac{x_q^{\dagger ai}v^\prime}{[p^2-(M_{q}^a)^2]^2}\,, \\
      (V^\dagger_{qL})^{aA}
      \frac{\bar{M}_{q}^{A}}{[p^2-(\bar{M}_{q}^{A})^2]^3}
      (V_{qR})^{Ai}
      &\rightarrow&
      \frac{x_q^{\dagger ai}v^\prime}{[p^2-(M_{q}^a)^2]^3}\,.
    \end{eqnarray}
It is found that these radiative corrections to the $\bar\theta$ parameter vanish. 
For example, $\delta\theta|_A$ is given as 
    \begin{eqnarray}
      \delta \theta|_A&=&
    {\rm Im\,Tr}\left(A^a_uA^b_d\right)f\left[(M^a_{u})^2,  (M^b_{d})^2, m_{\varphi'}^2 \right] \nonumber \\
      &=&
      \frac12 
      {\rm Im\,Tr}\left([A^a_u,A^b_d]\right)f\left[(M^a_{u})^2,  (M^b_{d})^2, m_{\varphi'}^2 \right] =0\,,
      \label{eq:LOvanish}
    \end{eqnarray}
    where $f[(M_u^a)^{2},  (M_d^b)^2, m_{\varphi'}^2]$ is the real function, and the  Hermitian property of the matrix $A_q^a$ is used.
    The same conclusions are applicable to $\delta \theta|_B$ and $\delta \theta|_C$ at this order.

 Now we showed that the charged NG boson contribution to the $\bar\theta$ parameter at two-loop level vanishes at the leading order of $v^\prime$ ($n=1$ in Eq.~(\ref{HDO})). It comes from the fact that the contributions are proportional to the fourth power of $x_{u/d}$. We have also checked that the contributions of the sixth power of $x_{u/d}$, corresponding to $\mathcal{O}(v^{\prime 4}/(M_q^a)^4)$ contributions, also vanish. The contributions are derived from the above formulae with the mass-insertion approximation,
    \begin{eqnarray}
      (V^\dagger_{qL})^{aA}
      \frac{i \bar{M}_{q}^{A}}{p^2-(\bar{M}_{q}^{A})^2}
      (V_{qR})^{Ai}
      &\rightarrow&
      i \frac{x_q^{\dagger ai}v^\prime}{p^2-(M_{q}^a)^2}\nonumber\\
      &&
      +i \frac{1}{p^2-(M_{q}^a)^2}(x_{q}^{\dagger}x_q)^{ab}v^{\prime2}
      \frac{x_q^{\dagger bi}v^\prime}{p^2-(M_{q}^b)^2}\,,\\
      (V^\dagger_{qL})^{aA}
      \frac{i \bar{M}_{q}^{A}}{[p^2-(\bar{M}_{q}^{A})^2]^2}
      (V_{qR})^{Ai}
      &\rightarrow&
      i \frac{x_q^{\dagger ai}v^\prime}{[p^2-(M_{q}^a)^2]^2}\nonumber\\
      &&
      +i \frac{1}{p^2-(M_{q}^a)^2}(x_q^\dagger x_q)^{ab}v^{\prime2}
      \frac{x_q^{\dagger bi}v^\prime}{[p^2-(M_{q}^b)^2]^2}\nonumber\\
      &&
      +i \frac{1}{[p^2-(M_{q}^a)^2]^2}(x_q^\dagger x_q)^{ab}v^{\prime2}
      \frac{x_q^{\dagger bi}v^\prime}{p^2-(M_{q}^b)^2}\,,\\
      (V^\dagger_{qL})^{aA}
      \frac{i\bar{M}_{q}^{A}}{[p^2-(\bar{M}_{q}^{A})^2]^3}
      (V_{qR})^{Ai}
      &\rightarrow&
      i\frac{x_q^{\dagger ai}v^\prime}{[p^2-(M_{q}^a)^2]^3}\nonumber\\
      &&
      +i\frac{1}{p^2-(M_{q}^a)^2}(x_q^\dagger x_q)^{ab}v^{\prime2}
      \frac{x_q^{\dagger bi}v^\prime}{[p^2-(M_{q}^b)^2]^3}\nonumber\\
      &&
      +i\frac{1}{[p^2-(M_{q}^a)^2]^2}(x_q^\dagger x_q)^{ab}v^{\prime2}
      \frac{x_q^{\dagger bi}v^\prime}{[p^2-(M_{q}^b)^2]^2}\nonumber\\
      &&
      +i\frac{1}{[p^2-(M_{q}^a)^2]^3}(x_q^\dagger x_q)^{ab}v^{\prime2}
      \frac{x_q^{\dagger bi}v^\prime}{p^2-(M_{q}^b)^2}\,.
    \end{eqnarray}
   Each first term is the aforementioned leading contribution, which vanishes (see Eq.~\eqref{eq:LOvanish}).
    The next-to-leading contributions are 
    \begin{align}
      \delta \theta|_A \simeq&  \frac1{8\pi^2}
      {\rm Im}\left[(A_u^{a})^{ij} (A_u^{b})^{jk} (A_d^c)^{ki}\right] \nonumber \\
     & \times  \left\{
      v^{\prime4}
      I_{(1,1;3)}\left[ (M^a_{u})^2, (M^b_{u})^2 ; (M^c_{d})^2 ;m^2_{\varphi^\prime} \right]
      -\frac{v^{\prime4}}{(M^c_{d})^2}
      B_{(1,1)}\left[0, (M^a_{u})^2, (M^b_{u})^2 ;m^2_{\varphi^\prime}\right]\right\}\nonumber\\
      &+\frac1{8\pi^2}
        {\rm Im}\left[(A_u^{a})^{ij} (A_d^{b})^{jk} (A_d^c)^{ki}\right]
        v^{\prime4}\left(I_{(1;1,3)}+I_{(1;2,2)}+I_{(1;3,1)}\right)\left[(M^a_{u})^2 ; (M^b_{d})^2 ,(M^c_{d})^2 ;m^2_{\varphi^\prime}\right]\,,
    \\
    \delta \theta|_B\simeq& \frac1{8\pi^2}
      {\rm Im}\left[(A_u^{a})^{ij} (A_d^{b})^{jk} (A_d^c)^{ki}\right]
      \nonumber \\
    & \times   \left\{
      v^{\prime4}
      I_{(3;1,1)}\left[ (M^a_{u})^2 ; (M^b_{d})^2 , (M^c_{d})^2 ;m^2_{\varphi^\prime} \right] -\frac{v^{\prime4}}{(M^a_{u})^2}
      B_{(1,1)}\left[0, (M^b_{d})^2 , (M^c_{d})^2 ; m^2_{\varphi^\prime} \right]\right\}\nonumber\\
      &+\frac1{8\pi^2}
        {\rm Im}\left[(A_u^{a})^{ij} (A_u^{b})^{jk} (A_d^c)^{ki}\right]
        v^{\prime4}\left(I_{(1,3;1)}+I_{(2,2;1)}+I_{(3,1;1)}\right)\left[ (M^a_{u})^2 , (M^b_{u})^2 ; (M^c_{d})^2 ; m^2_{\varphi^\prime} \right]\,,
   \\
        \delta\theta|_C\simeq& \frac{1}{8\pi^2}
    {\rm Im}
        \left[(A_u^{a})^{ij} (A_u^{b})^{jk} (A_d^c)^{ki}\right] v^{\prime4}
        \left({I}_{(1,2;2)}+{I}_{(2,1;2)}\right)
        \left[ (M_{u}^a)^2 , (M_{u}^b)^2 ; (M_{d}^c)^2 ; m^2_{\varphi^\prime} \right] \nonumber\\
     &+   \frac{1}{8\pi^2}
    {\rm Im}
        \left[(A_u^{a})^{ij} (A_d^{b})^{jk} (A_d^c)^{ki}\right] v^{\prime4}
        \left({I}_{(2:1,2)} +{I}_{(2:2,1)}\right)
        \left[ (M_{u}^a)^2 ; (M_{d}^b)^2 , (M_{d}^c)^2; m^2_{\varphi^\prime} \right] \,.
    \end{align} 
   These loop functions, which come from the mass-insertion approximation, are given in Appendix~\ref{app:loop}.  
   The sequences of masses connected by commas in the loop functions are introduced by the mass insertion.
   
   Again, it is found that
    these contributions are zero. For example, the first term in $\delta \theta|_A$ is given as
    \begin{eqnarray}
      \delta\theta|_A&=&
    {\rm Im\,Tr}\left(A^a_u A^b_u A^c_d \right)
    f \left[ (M^a_{u})^2 , (M^b_{u})^2 ; (M^c_{d})^2; m_{\varphi'}^2 \right]
    \nonumber\\
    &&+
    {\rm Im\,Tr}
    \left(A^a_u A^b_d A^c_d \right)
    g \left[ (M^a_{u})^2 ; (M^b_{d})^2 , (M^c_{d})^2; m_{\varphi'}^2 \right]
    \nonumber\\
    &=&
      \frac12 
    {\rm Im\,Tr}\left(A^c_d\, [A^a_u, A^b_u]\right) f\left[ (M^a_{u})^2 , (M^b_{u})^2 ; (M^c_{d})^2 ; m_{\varphi'}^2 \right] \nonumber \\
    &&
    +
    \frac12 
    {\rm Im\,Tr}\left(A^a_u\, [A^b_d, A^c_d]\right) 
    g \left[ (M^a_{u})^2 ; (M^b_{d})^2 , (M^c_{d})^2 ; m_{\varphi'}^2 \right]
    \nonumber \\
    &=& 0\,.
    \label{eq:sixth}
    \end{eqnarray}
    Here, above two real loop functions $f$ and $g$ are symmetric under exchanges of $(M^a_{u})^2 \leftrightarrow (M^b_{u})^2$  and $(M^b_{d})^2 \leftrightarrow (M^c_{d})^2$, respectively.  Then, the above equation vanishes, see Appendix~\ref{app:loop}. 
    The symmetry comes from the mass-insertion approximation.
Even if we include the higher-order contributions of $x_q$ in the mass-insertion approximation,
they still vanish since the loop function is real and symmetric for the exchange of the heavy fermion masses.

Now we found that the charged NG boson does not give a contribution to the $\bar\theta$ parameter at two-loop level. We also numerically checked this fact 
by using Eqs.~(\ref{two-loop-exact1A})--(\ref{two-loop-exact2}). 

The $W^{\prime\,\pm}$ contributions to the $\bar\theta$ parameter in the Feynman-'t Hooft gauge at two-loop level vanish. The Yukawa coupling dependence comes from only the mixing matrices, and then the leading contributions, which are proportional to the fourth power of $x_{u/d}$ at most, vanish.  The higher-order contributions, coming from mass-insertion approximation, also vanish due to the symmetry of heavy fermion masses in loop functions. 

A similar discussion is applicable for the other contribution, such as $Z^\prime$, $h^\prime$, and $\varphi^{\prime \, 0}$ at two-loop level. Then, we confirmed the two-loop contribution to the $\bar\theta$ parameter vanishes as far as $M_q\,\gsim \langle H^\prime \rangle\,\gg \langle H\rangle$.

\section{Non-vanishing contribution to QCD \texorpdfstring{\boldmath{$\theta$}}{theta} parameter in three-loop order}
\label{sec:3loop}

    In the previous section, we confirmed that the QCD $\bar\theta$ term is not generated in the two-loop level contribution, \ie, up to the fourth order of the Yukawa interaction $x_{q}$. 
    We also found that it is valid even if one considers the higher-order contributions of $x_q$ by using the mass-insertion approximation.
    In order to give non-vanishing contributions to the $\bar\theta$ parameter, the commutation relation $[A^a_q , A^b_q]$ must be nonzero, see Eq.~\eqref{eq:sixth}. 
    It implies that 
    non-vanishing contribution
    should be proportional to
    ${\rm Im\,Tr} (A^a_{q^\prime}\, [A^b_q, A^c_q] )$ for $q,q' = u$ and/or $d$ 
rather than ${\rm Im\,Tr} ( [A^b_q, A^c_q])$, 
    and the loop function has to 
    be asymmetric under exchange between 
    $(M_q^b)^2$ and $(M_q^c)^2$.
    Thus,   the contributions of the following form might be leading if 
    they are non-vanishing in the three-loop order,
    \begin{align}
        \delta \theta_{uuu} &\approx \frac{1}{(16\pi^2)^2} 
        \frac{v^{\prime 2}}{\widetilde{M}^2} \, {\rm Im\, Tr}\left( A^a_u \,[A^b_u , A^c_u] \right) f_{uuu}^{abc} \,,
        \label{eq:three-loop upper limit uuu} \\
        \delta \theta_{duu} &\approx  \frac{1}{(16\pi^2)^2} \frac{v^{\prime 2}}{\widetilde{M}^2} {\rm Im\,Tr}\left( A^a_d \,[A^b_u,A^c_u] \right) f_{duu}^{abc} \,,
        \label{eq:three-loop upper limit duu}
    \end{align}
    where $\widetilde{M}$ is the heaviest quark mass in the loop diagrams.
    Here, $f_{uuu}^{abc}$ is a dimensionless three-loop function which is totally antisymmetric under permutation of $(M_u^a)^2, (M_u^b)^2$ and $(M_u^c)^2$, while $f_{duu}^{abc}$ is antisymmetric under permutation of $ (M_u^b)^2$ and $ (M_u^c)^2$.
    Although the other types such as ${\rm Im\,Tr} (A_u^a\, [A_d^b , A_d^c])$ and ${\rm Im\,Tr} (A_d^a \,[A_d^b , A_d^c])$ can also contribute to the $\bar\theta$ parameter, they are suppressed by 
    the SM down-type quark masses, so we do not take them into account in this paper.

    We found that $\delta \theta_{uuu}$ in Eq.~\eqref{eq:three-loop upper limit uuu} is not generated from diagrams in the three-loop order in the minimal LR model. The corresponding three-loop diagrams do not have an asymmetric structure for three Dirac fermion masses when the internal scalar lines respect the $U(1)_{B-L}$. If the neutral scalar lines break the $U(1)_{B-L}$ (or the scalar lines pick  $v'$ using the four-point Higgs interaction), 
    the diagrams may have the asymmetric structure. However, in the case, the contribution the $\bar\theta$ parameter is proportional to $v^{\prime 4}/{\widetilde{M}}^4$ ($n=2$ in Eq.~\eqref{HDO}), not $v'^2/{\widetilde{M}}^2$. This situation is not changed even in the four-loop order. Thus, we conclude that $\delta \theta_{uuu}$ is not the leading contribution and   
    the largest non-vanishing contribution to the $\bar\theta$ parameter
    comes from $\delta \theta_{duu}$ in Eq.~(\ref{eq:three-loop upper limit duu})
    in the minimal LR symmetric model.

\subsection{Leading contribution: probability density function of \texorpdfstring{$\delta \theta_{duu}$}{delta theta duu}}

    \begin{figure}[t]
        \centering
        \includegraphics[width=0.6\linewidth]{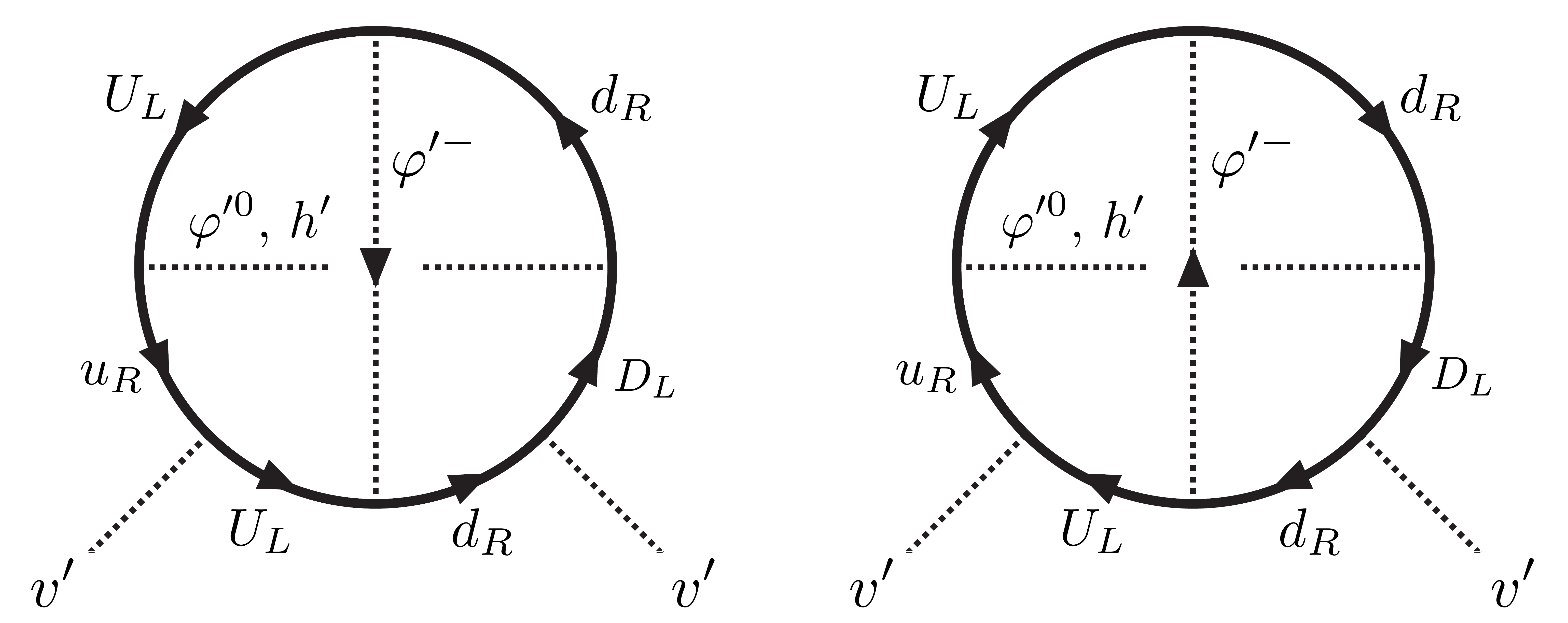}
        \caption{Diagrams that contribute to Eq.~(\ref{eq:three-loop upper limit duu}) and have asymmetric structure in the loop function. 
        The neutral scalar propagator corresponds to  the neutral NG boson $\varphi^{\prime\,0}$ and the neutral Higgs boson $h^{\prime}$.
        }
        \label{fig:three-loop duu}
    \end{figure}

In this section, we estimate the size of    
    $\delta \theta_{duu}$ in Eq.~(\ref{eq:three-loop upper limit duu}). We find that diagrams in Fig.~\ref{fig:three-loop duu} would provide the antisymmetric three-loop function and produce the non-vanishing  $\delta \theta_{duu}$.

When one considers a universal down-type vector-like quark mass $M_{d}^1 = M_{d}^2 =  M_{d}^3 \equiv \widetilde{M}_d$ for simplicity, 
     $\overline{\Phi}(\theta_{d3}, \theta_{d8})V_D$ in the down-type Yukawa $x_d$ in Eq.~\eqref{parametrization2}  become unphysical parameters,
     because these are removed by changing the basis of $D_{L/R}$. 
     Then, the contribution from $\delta \theta_{duu}$ is simplified as
    \begin{align}
            \delta \theta_{duu} \approx& \frac{1}{(16 \pi^2)^2} \frac{v^{\prime 2}}{\widetilde{M}^2} \frac{\widetilde{M}_d M_u^b M_u^c}{v^{\prime 3}} 
            \frac{m_d^i m_u^k \sqrt{m_u^j m_u^l}}{v^3}  \nonumber \\
            &\times {\rm Im\,Tr} \left[ V_{\rm CKM}^{\dagger ij} V_U^{\prime jb} V_U^{\prime \dagger bk} V_U^{\prime kc} V_U^{\prime \dagger cl} V_{\rm CKM}^{li}
            - {(b\leftrightarrow c)}
            \right] \tilde{f}_{duu}^{bc}\nonumber \\
            \approx& \frac{4}{(16 \pi^2)^2} \frac{v^{\prime 2}}{\widetilde{M}^2} \frac{\widetilde{M}_d M_u^b M_u^c}{v^{\prime 3}} \frac{m_b m_t^{\frac{3}{2}} \sqrt{m_c}}{v^3} \operatorname{Im} \left( V_{\rm CKM}^{\dagger 33} V_U^{\prime 3b} V_U^{\prime \dagger b3} V_U^{\prime 3c} V_U^{\prime \dagger c2} V_{\rm CKM}^{23} \right) \tilde{f}_{duu}^{bc}\,,
        \label{eq:three-loop duu}
    \end{align}
where
a three-loop function 
$\tilde{f}^{bc}_{duu}$ is antisymmetric under the permutation of $b$ and $c$, and
$\tilde{f}^{bc}_{duu} = f^{1bc}_{duu} = f^{2bc}_{duu} =f^{3bc}_{duu}$ for the universal down-type Dirac quark mass.
Here, $V_U^\prime \equiv \overline{\Phi} (\theta_{u3}, \theta_{u8}) V_U$.
A term proportional to $m_b m_t^2$ (corresponding to $i=3$ and $j=k=l=3$) vanishes by $\text{Im}(V_{\rm CKM}^{\dagger 33} V_U^{\prime 3b} V_U^{\prime \dagger b3} V_U^{\prime 3c} V_U^{\prime \dagger c3} V_{\rm CKM}^{33})=0$.
Although the above contribution ($i=3$, $j=k=3$ and $l=2$) is suppressed by $V_{\rm CKM}^{23} \simeq 0.04$, we found that it can provide a larger contribution than a term
proportional to $m_b m_t m_c$ ($i=3$, $j=l=3$ and $k=2$).

We considered the benchmark points where $\widetilde{M} = M_d^1 = M_d^2 = M_d^3 = M_u^1 = M_u^2 = 10^3 M_u^3$ and $\widetilde{M} = M_d^1 = M_d^2 = M_d^3 = M_u^1 = M_u^2 = 10^2 M_u^3$. 
Both benchmark points have the degenerate down-type vector-like quark masses and
the partially degenerate up-type masses.
This is because we would like to focus on the case that the hierarchy in the SM down-type quark masses is explained by not one in the down-type vector-like quark masses $M_d^a$ but one in the components of the Yukawa coupling $x_d^{ia}$ in the seesaw mechanism Eq.~(\ref{eq:seesaw}). This is motivated by the fact that the SM down-type quark masses have a moderate hierarchy compared to the up-type ones.

We estimate the size of the three-loop function as $v^{\prime 2}/\widetilde{M}^2  \tilde{f}_{duu}^{bc}$, where $\widetilde{M}$ is the heaviest quark mass. 
It is naively expected that when a loop function is made up of $\mathcal{O}(1)$ mass-ratio parameter, its size is maximized.
In this case a mass-ratio  $m_{\varphi'}^2/(M_u^3)^2$ can become $\mathcal{O}(1)$, so that the three-loop function would be maximized when $b$ or $c=3$.
Therefore, 
the leading contributions in Eq.~(\ref{eq:three-loop duu}) would be 
dominated by $(b,c)=(1,3)$ and $(2,3)$.

\begin{figure}[t]
\begin{center}
\begin{subfigure}{0.48 \textwidth}
\includegraphics[width = \textwidth]{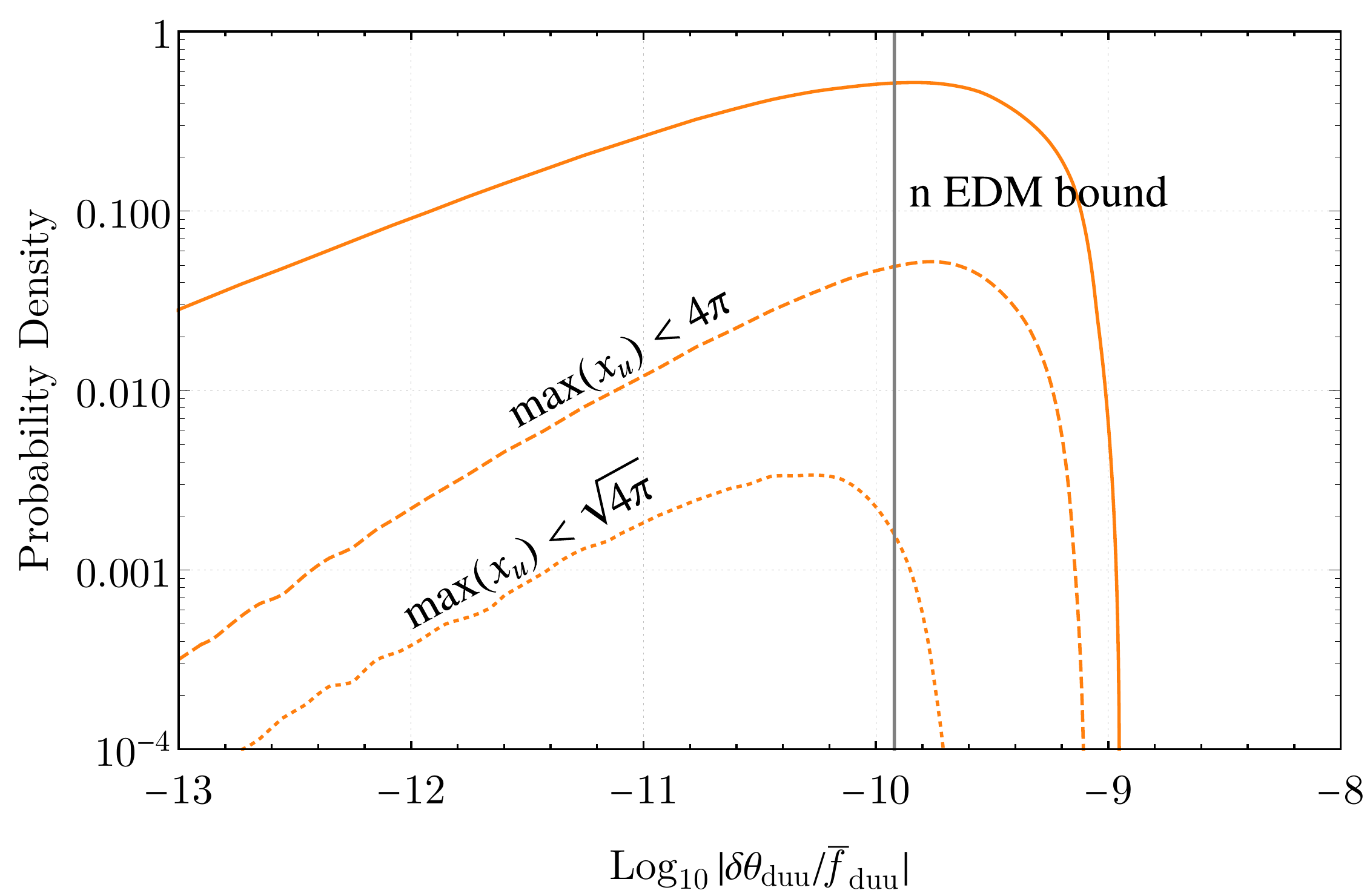}
\caption{$M_u^1 = 10^3 M_u^3$}
\label{fig:duu1}
\end{subfigure}
\quad 
\begin{subfigure}{0.48 \textwidth}
\includegraphics[width = \textwidth]{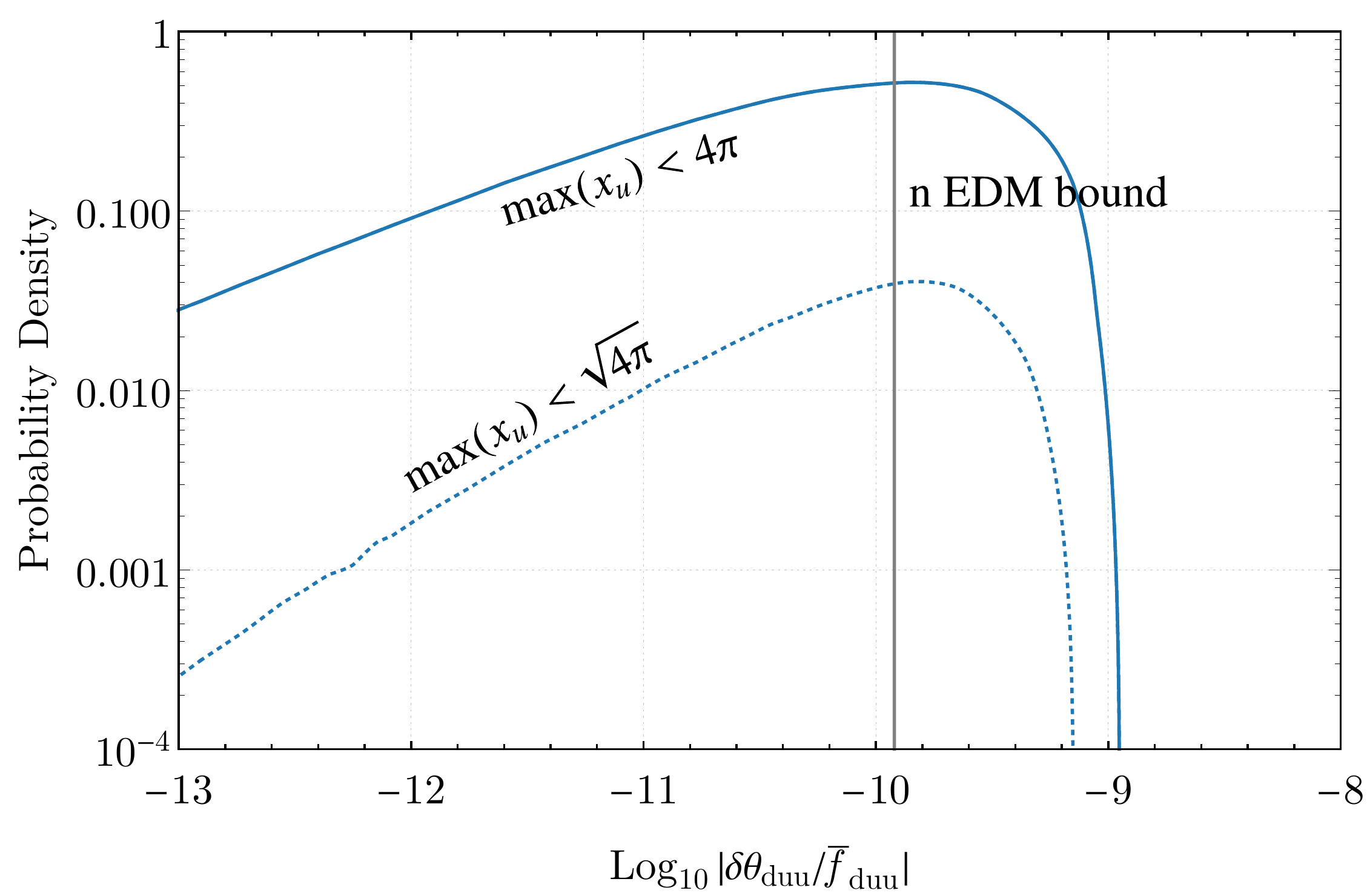}
\caption{$M_u^1 = 10^2 M_u^3$}
\label{fig:duu2}
\end{subfigure}
\caption{
The probability density functions (PDFs)  of $|\delta \theta_{duu}|$ in Eq.~\eqref{eq:three-loop upper limit duu}
with only $b=1$ and $c=3$ normalized by
a three-loop function
$|\bar{f}_{duu}^{13}|$ (solid line). 
We take $\widetilde{M} = \widetilde{M}_d = M_u^1 = 10^3 M_u^3$ in (a) and $\widetilde{M} = \widetilde{M}_d = M_u^1 = 10^2 M_u^3$ in (b).
After imposing the perturbative unitarity bound of $\textrm{max}(x_u) < 4 \pi \,(\sqrt{4 \pi})$, the PDFs are expected by the dashed (dotted) lines.
The areas of dashed and dotted lines are not normalized by one (see text).
Note that the solid and dashed lines overlap in (b).
 If $\bar{f}_{duu}^{13}=1$, 
 the right area of the vertical line is excluded  at 90\% CL by the neutron EDM measurement~\eqref{eq:constraint theta}.
}
\label{fig:duu}
\end{center}
\end{figure}
\begin{figure}[t]
\begin{center}
\begin{subfigure}{0.48 \textwidth}
\includegraphics[width = \textwidth]{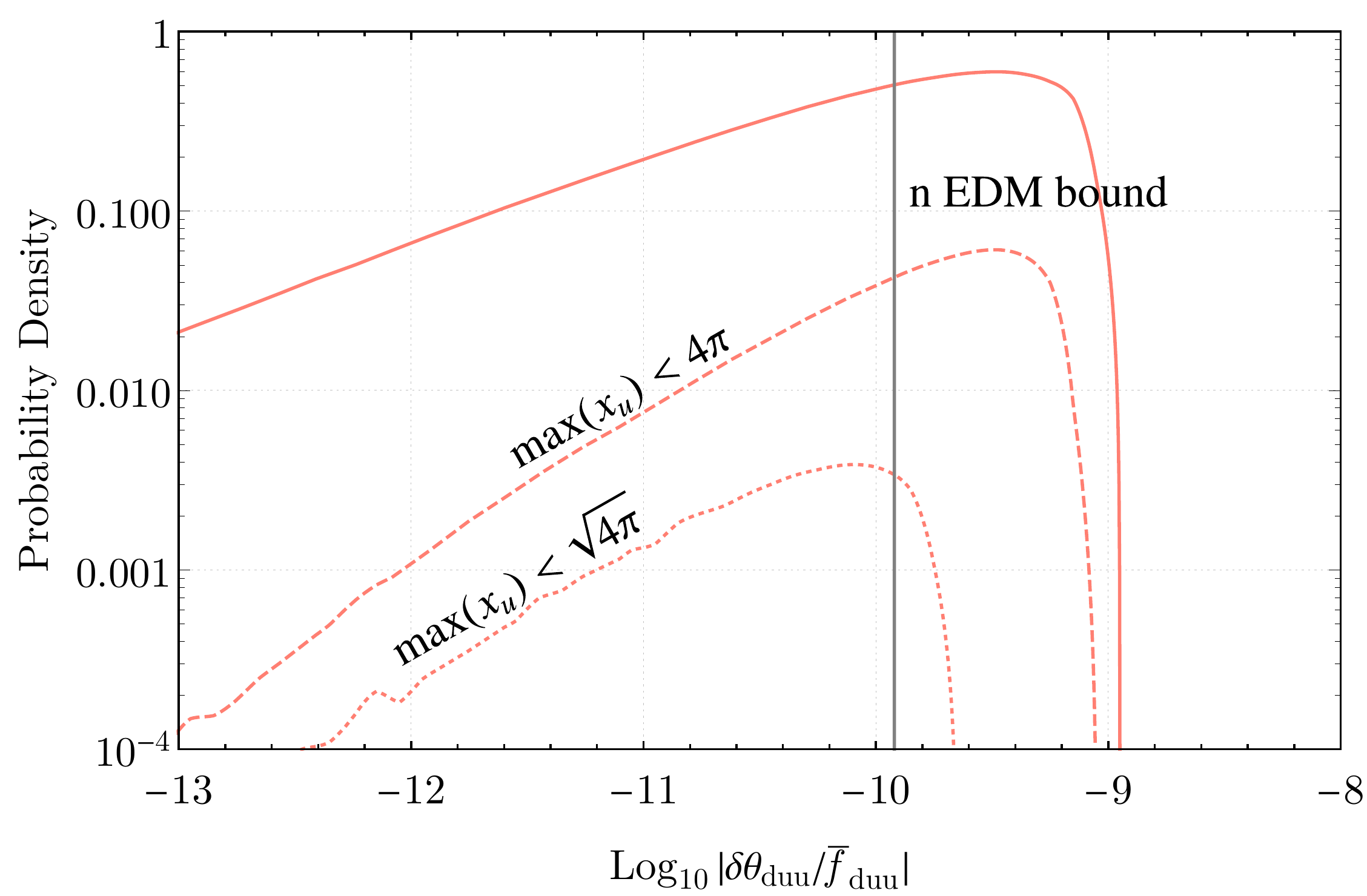}
\caption{$M_u^1  = M_u^2= 10^3 M_u^3$}
\label{fig:duu3}
\end{subfigure}
\quad 
\begin{subfigure}{0.48 \textwidth}
\includegraphics[width = \textwidth]{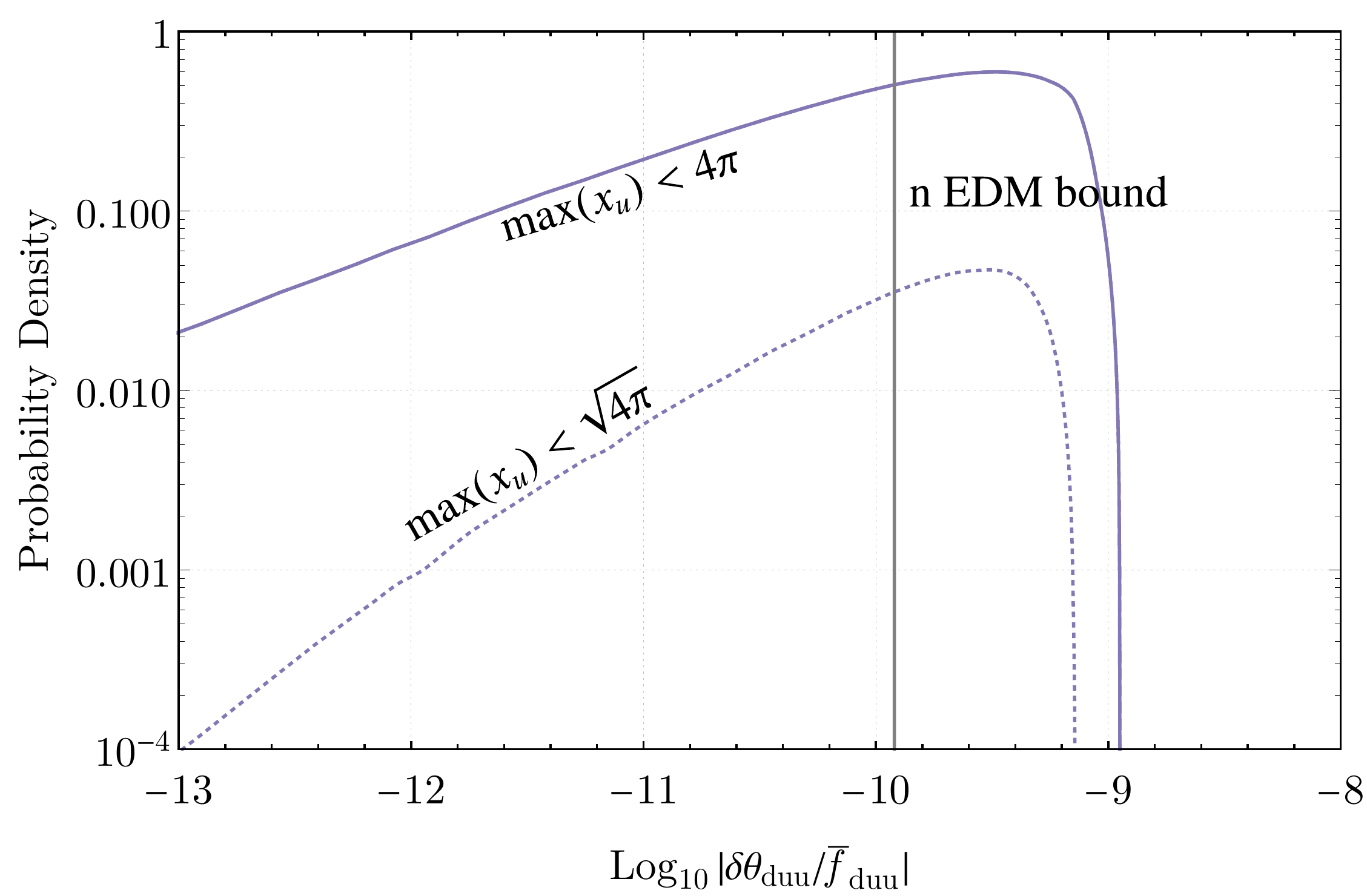}
\caption{$M_u^1 = M_u^2 =10^2 M_u^3$}
\label{fig:duu4}
\end{subfigure}
\caption{
The PDFs of $|\delta \theta_{duu}|$ in Eq.~\eqref{eq:three-loop upper limit duu}
with $b=1,\,2$ and $c=3$  normalized by $|\bar{f}_{duu}^{13}|=|\bar{f}_{duu}^{23}|$ (solid line). 
We take $\widetilde{M} = \widetilde{M}_d = M_u^1  = M_u^2= 10^3 M_u^3$ in (a) and $\widetilde{M} = \widetilde{M}_d = M_u^1  = M_u^2= 10^2 M_u^3$ in (b).
After imposing the perturbative unitarity bound of $\textrm{max}(x_u) < 4 \pi \,(\sqrt{4 \pi})$, the PDFs are expected by the dashed (dotted) lines.
Note that the solid and dashed lines overlap in (b).
 If $\bar{f}_{duu}^{13}=\bar{f}_{duu}^{23}=1$, 
 the right area of the vertical line is excluded  at 90\% CL by the neutron EDM measurement~\eqref{eq:constraint theta}.
}
\label{fig:duu1and2}
\end{center}
\end{figure}
In Figs.~\ref{fig:duu1} and \ref{fig:duu2}, we show the probability density functions (PDFs) of the absolute value of $\delta \theta_{duu}$ in Eq.~\eqref{eq:three-loop upper limit duu}
with $b=1$ and $c=3$ (also included $b=3$ and $c=1$), by the solid lines.
The total integrals of the solid line PDFs are $1$ in all plots.
 When $ M_u^1 \gg  M_u^2$ (reflecting the fact that $m_u \ll m_c)$, this contribution would be dominant in the non-vanishing $\delta \theta_{duu}$, see Eq.~\eqref{eq:three-loop duu}.
In the PDFs, 
the $\delta \theta_{duu}$ is normalized by $|\bar{f}_{duu}^{13}|$ where we define 
$\bar{f}_{duu}^{13} = \tilde{f}^{13}_{duu} -
 \tilde{f}^{31}_{duu} = 2 \tilde{f}^{13}_{duu}
 $.
 In Fig.~\ref{fig:duu1} and \ref{fig:duu2}, we take $\widetilde{M} = \widetilde{M}_d = M_u^1 = 10^3 M_u^3$ and $\widetilde{M} = \widetilde{M}_d = M_u^1 = 10^2 M_u^3$, respectively, with $v' = M_u^3$, by solid lines,  whose hierarchy is concerned with the top quark masses in the SM and the seesaw mechanism in Eq.~(\ref{eq:seesaw}).
 After fixing the ratios of the mass parameters, and assuming that all light quark masses and the CKM components reproduce the SM one, 
 the remaining free parameters are the three rotation angles and one $CP$-violating phase in the $V_U$ matrix and 
 two additional $CP$-violating phases in $\overline{\Phi}(\theta_{u3}, \theta_{u8})$ in Eq.~\eqref{parametrization2}.
 In order to obtain the PDFs, 
 we varied these six angles from $0$ to $2\pi$ with equal probability. This provides the solid line PDFs
 in Fig.~\ref{fig:duu1} and \ref{fig:duu2}.

 Although the radiative $\bar\theta$ parameter should be renormalization scale invariant,\footnote{%
 Recently, there is a discussion of a perturbative running of a renormalized $\theta$ parameter within renormalizable theories \cite{Valenti:2022uii}.
In our setup, the bare $\theta$ parameter (hence the renormalized one) is forbidden.} 
in order to investigate the perturbative unitarity bounds for the Yukawa interactions, we used the SM (running) quark masses at $\mu=1\,$TeV.\footnote{%
We take 
$m_t = 143 \, {\rm GeV}, m_b = 2.41 \, {\rm GeV}, m_c = 528 \, {\rm MeV}$, and $m_s = 45.4 \, {\rm MeV}$  at $\mu=1$~TeV \cite{Chetyrkin:2000yt}. 
} 
Furthermore, by assuming  $\bar{f}_{duu}^{13}=1$, 
the vertical line in the figures stands for 
the experimental upper bound from the neutron EDM measurement in Eq.~(\ref{eq:constraint theta}) and the right area is excluded at 90\% CL.

Given the fixed values of $\overline{\Phi}(\theta_{u3},\theta_{u8}) V_U$, 
one can obtain the eigenvalues of the up-type Yukawa matrix $x_u$.
In the PDF analysis, 
 we find that the eigenvalues of the Yukawa matrix $x_u$ can be larger than $O(1)$ easily depending on $\overline{\Phi}(\theta_{u3},\theta_{u8}) V_U$. 
 Therefore, 
 we impose the maximal  eigenvalue to be smaller than $4 \pi$ or $\sqrt{4 \pi}$  for the dashed  or dotted lines, respectively, as the perturbative unitarity bound. 
When the maximal eigenvalue exceeds the unitary bound, we discarded these points in the PDFs.
 In order to display the reduction in statistics as a result of setting the perturbative bound, we do not normalize the dashed and dotted line PDFs by $1$, and hence their total integration is less than $1$.

     \begin{table}[t]
        \centering
        \newcommand{\bhline}[1]{\noalign{\hrule height #1}}
        \renewcommand{\arraystretch}{1.5}
        \rowcolors{2}{gray!15}{white}
        \addtolength{\tabcolsep}{5pt} 
        \begin{tabular}{c c c}
                    \bhline{1 pt}        
                     &      $M_u^3 / M_u^1 = 10^{-3}$
                     &  $M_u^3 / M_u^1 = 10^{-2} $\\
                    \hline 
                     $\textrm{max} (x_u) <4 \pi$
                     &  $42.1 \% \,(58.9\%)$ & $32.2 \% \,(46.2\%)$ \\
                     $\textrm{max} (x_u) < \sqrt{4 \pi}$
                     & $3.65 \% \,(10.6\%)$ & $38.8 \% \,(55.9\%)$ \\
                    \bhline{1 pt}
                \end{tabular}
        \addtolength{\tabcolsep}{-5pt} 
        \caption{
        Ratios of excluded parameter regions for the parameter sets in Fig.~\ref{fig:duu} 
        from the current neutron EDM measurement, 
        under an assumption of  $\bar{f}_{duu}^{13}=1$.
        The numbers in parentheses are those in
        Fig.~\ref{fig:duu1and2}
         under an assumption of  $\bar{f}_{duu}^{13}=\bar{f}_{duu}^{23}=1$.
        The parameter sets are restricted by the perturbative bound of the Yukawa coupling $x_u$ as $4 \pi$ or $\sqrt{4\pi}$. Here, $\widetilde{M}=\widetilde{M}_d =M_u^1=M_u^2$.
        }
        \label{tab:duu}
    \end{table}
  
The ratios of the excluded parameter regions by the neutron EDM measurement in each of the PDFs are shown in Table~\ref{tab:duu} under an assumption of $\bar{f}_{duu}^{13}=1$.
Here, the ratios of the excluded parameter regions are obtained by comparing the areas of the dashed or dotted line PDFs; right area of the vertical line over the total area where the unitarity bound is imposed.
We found that some fractions of parameter regions have already been excluded even if the perturbative bound is imposed. 
In the case of the perturbative bound of  $4 \pi$ to the eigenvalues of $x_u$ matrix, 
about 30--40\% of the whole parameter region is already excluded.
On the other hand, in the case of $\sqrt{4 \pi}$, the dependence of $M_u^3$ appears obviously. In particular, only 3.65\% is excluded in the case of $M_u^1 = 10^3 M_u^3$.  
These results show that this model is sensitive to the current bound from the neutron EDM experiments and has a possibility to be explored by future improvement of the experiments.

Moreover, in Figs.~\ref{fig:duu3} and \ref{fig:duu4},
we show the PDFs of the absolute value of $\delta \theta_{duu}$ with $(b,c)=(1,3)$ plus  $(2,3)$ (also included $(3,1)$ and $(3,2)$), with assuming $M_u^1 = M_u^2$ which should be a somewhat aggressive parameter choice in light of  $m_u \ll m_c$.
We observed that the PDFs in Fig.~\ref{fig:duu1and2} are slightly larger than the PDFs in Fig.~\ref{fig:duu}.
The ratios of excluded parameter regions by the neutron EDM measurement are shown in
Table~\ref{tab:duu} at the numbers in parentheses.

Above estimation of the leading contribution to the radiative $\bar\theta$ parameter is numerically consistent with the latest analysis Eq.~(28) of Ref.~\cite{deVries:2021pzl}, where the ordinary calculation method is used. 
In this paper, 
we showed for the first time that an effect from the perturbative unitarity bound is important and it reduces the radiative corrections to $\bar\theta$.

\subsection{Case for the GIM by universal vector-like mass}

    Next,  
    we investigate a case that 
    all Dirac quark masses are degenerate as $\widetilde{M}=  M_u^a = M_d^a$.
    In this case, the GIM-like mechanism occurs and then the CKM matrix becomes a unique source of the $CP$-violating phase.
    As a consequence, 
    the radiative $\bar\theta$ parameter would be significantly suppressed, and it is expected as the minimum value of the $\bar\theta$ parameter in the minimal LR symmetric model. 
    The induced $\bar\theta$ parameter should be proportional to the Jarlskog invariant of the CKM matrix $J_{\rm CKM}$, which is given by 
    $ \text{Im}( V_{\rm CKM}^{ij} V_{\rm CKM}^{\dag jk}   V_{\rm CKM}^{kl}  V_{\rm CKM}^{\dag li} )  = J_{\rm CKM} \sum_{m,n} \epsilon_{ikm} \epsilon_{jln}$
    \cite{Jarlskog:1985ht}
    and 
    $J_{\rm CKM} = ( 3.08 {}^{+0.15}_{-0.13} ) \times 10^{-5}$~\cite{ParticleDataGroup:2022pth}.

    If the $\bar\theta$ parameter is induced at  three-loop level,
    the contribution would be given 
    as 
    \begin{align}
            \delta \theta_{\rm minimum} &\approx \frac{1}{\left(16 \pi^2\right)^2} \frac{v^{\prime 8}}{\widetilde{M}^8} \operatorname{Im} \operatorname{Tr}\left(A_u^2 A_d^2 A_u A_d\right) {f_{\rm CKM}} \nonumber \\
            &=\frac{1}{\left(16 \pi^2\right)^2} \frac{v^{\prime 2}}{\widetilde{M}^2} J_{\rm CKM} {f_{\rm CKM}}  \nonumber\\
            & \quad \times \frac{1}{v^6} \left(m_t-m_c\right)\left(m_t-m_u\right)\left(m_c-m_u\right)\left(m_b-m_s\right)\left(m_b-m_d\right)\left(m_s-m_d\right), 
         \nonumber  \\
            &\simeq 1 \times 10^{-{19}} \frac{v^{\prime 2}}{\widetilde{M}^2} {f_{\rm CKM}} \,,
        \label{eq:min theta}
    \end{align}
    where $f_{\rm CKM}$ is a dimensionless loop function of $O(1)$. It is assumed that in the three-loop diagrams, two scalar bosons are exchanged inside the fermion loop, and the other scalar lines are replaced by $v^\prime$.  The above contribution is proportional to $v^{\prime 8}$.
    The perturbativity of the top Yukawa requires $\widetilde{M}\simeq  {v'}$.
    Then, $\delta \theta_{\rm minimum}$ is expected as at most $10^{-{19}}$ or smaller when  all vector-like quark masses are degenerate. 
    This size is comparable to or smaller than the CKM phase contribution to the $\bar\theta$ parameter in the SM at four-loop level, evaluated in Ref.~\cite{Khriplovich:1985jr} (though the top quark had been assumed to be lighter than the $W$ boson). This is because the quark mass suppression of the contribution in the SM  is milder than Eq.~\eqref{eq:min theta}.
    In addition, the neutron EDM induced   by the CKM phase via long-distance hadronic contributions in the SM \cite{Dar:2000tn, Mannel:2012qk} is much larger than by the contribution to the $\bar\theta$ parameter in the above benchmark point.    Then, a more precise evaluation of  the $\bar\theta$ parameter in the benchmark point  is too academic and beyond our scope.

\section{Conclusions and discussion}
\label{sec:conclusions}

When the QCD axion is absent,
one has to solve the strong $CP$ problem by an additional discrete symmetry.
The extended parity with the LR gauge symmetry
can solve it with generating the SM as the low-energy theory. 
However, it is known that the radiative $\bar\theta$ parameter is induced from the soft symmetry breaking of the parity.
In this paper, we first formulated a novel method of direct loop-diagrammatic calculation
of the radiative $\bar \theta$ parameter by using Fock-Schwinger gauge. 
This approach should be more robust than the 
the ordinary calculation method 
based on the chiral rotations.
By using the Fock-Schwinger gauge method,
we confirmed a seminal result that 
two-loop level $\bar\theta$ vanishes  completely 
in the minimal LR symmetric model.

Furthermore, 
we estimated the size of the leading contributions to the  non-vanishing radiative $\bar\theta$ parameter   at three-loop level.
We derive the parameterization by the physical parameters  based on the seesaw mechanism in the LR symmetric model, and we obtained the probability density functions
of the radiative $\bar\theta$ parameter by varying all free but physical parameters. 
Here, we also investigated the impact of the perturbative unitarity conditions for the LR symmetric Yukawa matrices.
It is found that the resultant $\bar\theta$ parameters are partially excluded by
the current neutron EDM bound. 
It implies that this model has a possibility to be explored by future improvement of the experiments.
One should note that the minimal LR symmetric model predicts that all hadronic EDMs are dominated by the radiative $\bar\theta$ parameter. Therefore, this model can predict a distinctive and non-vanishing correlation between the neutron, proton, nuclei (${}^2$H, ${}^3$He),  diamagnetic atoms (Hg, Ra), paramagnetic atoms and molecules (YbF, HfF, ThO) EDMs \cite{deVries:2021pzl,Dekens:2014jka,deVries:2018mgf,deVries:2021sxz}.

The large mass-scale difference between $v$ and $v'$ can be explained by a Higgs parity mechanism that predicts 
$\lambda_{\rm SM}(\mu = v^\prime)\simeq 0$ and $v' = \mathcal{O}(10^{10})$\,GeV  \cite{Hall:2018let,Dunsky:2019api,Hall:2019qwx,Dunsky:2019upk}. 
Since the above estimation of the radiative $\bar\theta$ parameter is insensitive to the mass scale itself of the right-handed sector, one could predict the neutron EDM for the case of  $v' = \mathcal{O}(10^{10})$\,GeV.

We also comment on the radiative $\bar{\theta}$ parameter in the spontaneous $CP$ violation (Nelson-Barr) model \cite{Nelson:1983zb,Barr:1984qx,Barr:1984fh}, which is another model that can explain the strong $CP$ problem by the discrete symmetry. In Ref.~\cite{Valenti:2021rdu}, the radiative $\bar\theta$ parameter has been investigated in detail. It is found that although the reducible $\bar\theta$, which comes from quartic couplings of scalar fields responsible for the spontaneous $CP$ violation with the SM Higgs one, is induced at two-loop level, it can be numerically neglected if the couplings are small enough. On the other hand, the irreducible $\bar\theta$, which is related to the CKM phase, is induced at three-loop level and is safely below the current experimental bounds. The loop-diagrammatic approach we proposed would provide a more robust estimation if possible.

It would be an interesting direction 
to investigate correlations between the radiative $\bar \theta$ parameter and other (flavor) observables in 
the minimal LR symmetric model:
The lepton flavor universality violation in $B \to D^{(\ast)} l \nu$ ($R(D^{(\ast)}) $ anomaly) (the recent review  \cite{HFLAV:2022pwe,Iguro:2022yzr}), the Cabibbo angle anomaly (the recent review \cite{Crivellin:2022rhw}) and the $W$-boson mass anomaly \cite{CDF:2022hxs} could  be explained \cite{Babu:2018vrl,Babu:2022jbn,Dcruz:2023mvf}.
Furthermore, the investigation of a correlation with electroweak-like baryogenesis in the right-handed sector should be an attractive prospect \cite{Fujikura:2021abj,Harigaya:2022wzt}.

\acknowledgments

We would like to acknowledge  Keisuke Harigaya  for a collaboration at the early stage.
This work is supported by the JSPS Grant-in-Aid for Scientific Research Grant No.\,20H01895 (J.H.),
No.\,21K03572 (J.H.) and 
for Early-Career Scientists Grant No.\,19K14706 (T.K.).
The work of J.H.\ is also supported by 
World Premier International Research Center Initiative (WPI Initiative), MEXT, Japan.
This work is also supported by 
JSPS Core-to-Core Program Grant No.\,JPJSCCA20200002. 
This work was financially supported by JST SPRING, Grant Number JPMJSP2125. The 
author (A.Y.) would like to take this opportunity to thank the ``Interdisciplinary Frontier 
Next-Generation Researcher Program of the Tokai Higher Education and Research System.''

\appendix


\section{Loop functions}
\label{app:loop}

    We define the loop functions in this section. The two-loop functions used in this paper are given as
    \begin{align}
      &I_{(n_1,\cdots;m_1,\cdots)}(x_1,\cdots;x_2,\cdots;x_3) \nonumber \\
      &\quad =(16\pi^2\mu^{2\epsilon})^2\int \frac{d^dp}{(2\pi)^{d}}\frac{d^dq}{(2\pi)^{d}}
       \frac{1}{[p^2-x_1]^{n_1}\cdots[q^2-x_2]^{m_1}\cdots[(p+q)^2-x_3]}\,,
    \end{align}  
    where the dimensional regularization is used on $d=4-2\epsilon$ dimension and $\mu$ is the renormalization scale.
    The functions can also be derived by derivative or finite difference of $I(x_1;x_2;x_3)$ ($\equiv I_{(1;1)}(x_1;x_2;x_3)$)
    as
    \begin{eqnarray}
      I_{(n;m)}(x_1;x_2;x_3)&=&\frac{1}{(n-1)!(m-1)!}
      \frac{d^{n-1}}{dx_1^{n-1}}\frac{d^{m-1}}{dx_2^{m-1}}
      I(x_1;x_2;x_3)\,,\\
      I_{(1,1;1)}(x_1,x_1^\prime;x_2;x_3)&=&\frac{1}{x_1-x_1^\prime}\left[
      I(x_1;x_2;x_3)-I(x_1^\prime;x_2;x_3)\right]\,,
    \end{eqnarray}  
    where
    \begin{eqnarray}
      I(x_1;x_2;x_3)&\equiv &I_{(1;1)}(x_1;x_2;x_3)\nonumber\\
      &=&(16\pi^2\mu^{2\epsilon})^2\int \frac{d^dp d^dq}{(2\pi)^{2d}}
        \frac{1}{[p^2-x_1][q^2-x_2][(p+q)^2-x_3]}\,.
    \end{eqnarray}
    The explicit form of
    $I(x_1;x_2;x_3)$ is given as
    \begin{eqnarray}
      I(x_1;x_2;x_3)&=&
    \bar{I}_\epsilon(x_1;x_2;x_3)+\bar{I}(x_1;x_2;x_3)\,,
    \end{eqnarray}
    with the UV divergent part
    \begin{align}
    \bar{I}_\epsilon(x_1;x_2;x_3)=-\sum_{i=1,2,3}x_i\left[
      \frac{1}{2\epsilon^2}-\frac{1}{\epsilon}\left(\log\frac{x_i}{Q^2}-\frac32\right) + \left(1+\frac{\pi^2}{12}-\log\frac{x_i}{Q^2}+\frac12 \log^2\frac{x_i}{Q^2}\right)\right]\,,
    \end{align}
    while the finite part 
     \begin{align}
     & \bar{I}(x_1;x_2;x_3)
      =
      -\frac12\biggl[(-x_1+x_2+x_3)\log\frac{x_2}{Q^2}\log\frac{x_3}{Q^2}
    +(x_1-x_2+x_3)\log\frac{x_1}{Q^2}\log\frac{x_3}{Q^2}
    \nonumber\\
    &\quad 
    +(x_1+x_2-x_3)\log\frac{x_1}{Q^2}\log\frac{x_2}{Q^2}
    -4\left(x_1\log\frac{x_1}{Q^2}
    +x_2\log\frac{x_2}{Q^2}
    +x_3\log\frac{x_3}{Q^2}\right)
    \nonumber\\
    &\quad 
    +5(x_1+x_2+x_3)+\xi(x_1,x_2,x_3)\biggr]\,,
   \\
   & \xi(x_1,x_2,x_3)=
    R\left[2 \log\left(\frac{x_3+x_1-x_2-R}{2 x_3}\right)\log\left(\frac{x_3-x_1+x_2-R}{2 x_3}\right)
    -\log\frac{x_1}{x_3}\log\frac{x_2}{x_3}
    \right.\nonumber\\
    &
   \quad  \left.
    -2{\rm Li}_2\left(\frac{x_3+x_1-x_2-R}{2 x_3}\right)
    -2{\rm Li}_2\left(\frac{x_3-x_1+x_2-R}{2 x_3}\right)
    +\frac{\pi^2}{3}
    \right]\,,
    \end{align}
    where $R=\sqrt{x_1^2+x_2^2+x_3^3-2 x_1 x_2-2x_2 x_3-2 x_3 x_1}$ and
    $Q^2\equiv4\pi \mu^2 {\rm e}^{-\gamma_E}$ \cite{Ford:1992pn,Espinosa:2000df,Martin:2001vx}. 
    Note that although the last terms of $\bar{I}_\epsilon$ are UV finite, 
    they do not affect any physical quantity \cite{Martin:2018emo}. 
    These terms are suppressed by $\mathcal{O}(\epsilon)$ at one-loop level, while they are uplifted to  $\mathcal{O}(\epsilon^0)$ at two-loop level.

    \subsection{\texorpdfstring{$I_{(1;3)}$}{I{(1;3)}} and \texorpdfstring{$I_{(3;1)}$}{I{(3;1)}}}
The two-loop function $I_{(3;1)}$ is also given by the two-point one-loop function $B_0$ as 
    \begin{eqnarray}
      I_{(3;1)}(x_1;x_2;x_3)&=&-(16\pi^2\mu^{2\epsilon})
      \int\frac{d^dp}{i(2\pi)^d}\frac{1}{(p^2-x_1)^3}
      B_0(p^2,x_2,x_3)\,,
    \end{eqnarray}
    where
    \begin{align}
      B_0(p^2,x_2,x_3) &\equiv  (16\pi^2\mu^{2\epsilon}) \int \frac{d^d q}{i (2\pi)^d} \frac{1}{[q^2 - x_2][(p+q)^2-x_3]} \nonumber \\
      &=\frac1{\epsilon}-F_0(p^2,x_2,x_3)\,,\\
      F_0(p^2,x_2,x_3)&=\int^1_0dz \log\frac{-z(1-z)p^2+z x_2+(1-z)x_3}{Q^2}\,.  
    \end{align}
         Similarly, $\bar{I}_{(1;3)}$ is obtained by  replacements of $x_1 \leftrightarrow x_2$ in the above equations of $\bar{I}_{(3;1)}$.  
         
    The finite part of ${I}_{(3;1)}$ ($=\bar{I}_{(3;1)}$) is given as 
    \begin{eqnarray}
      \bar{I}_{(3;1)}(x_1;x_2;x_3)&=&(16\pi^2\mu^{2\epsilon})\int\frac{d^dp}{i(2\pi)^d}
        \frac{1}{(p^2-x_1)^3} F_0(p^2,x_2,x_3)\,,
    \end{eqnarray}
while the divergent part is
\begin{align}
    \bar{I}_{\epsilon{(3;1)}}(x_1;x_2;x_3)= \frac{1}{2  x_1} \left(\frac{1}{\epsilon}- \log \frac{x_1}{Q^2}\right) \,.
    \label{eq:UVdiv}
\end{align}

    In the case of $x_1\ll x_3$,
    $\bar{I}_{(3;1)}$ is given as 
    \begin{eqnarray}
      \bar{I}_{(3;1)}(x_1;x_2;x_3)&\simeq &-\frac{1}{2 x_1}F_0(0,x_2,x_3)\,,
    \end{eqnarray}
    where
    \begin{eqnarray}
      F_0(0,x_2,x_3)&=&\int^1_0dz \log\frac{z x_2+(1-z) x_3}{Q^2}\nonumber\\
      &=&-\frac{x_2-x_3 -x_2\log(x_2/Q^2)+x_3\log(x_3/Q^2)}{x_2-x_3}\,.
    \end{eqnarray}
On the other hand, in the case of $x_2 \ll x_3$, $\bar{I}_{(3;1)}$ is given as
    \begin{eqnarray}
    \bar{I}_{(3;1)}(x_1;x_2;x_3)&\simeq& 
    \frac{1}{2 x_1} \frac{x_1 - x_3 - x_1 \log(x_1/Q^2) + x_3 \log(x_3/Q^2)}{x_1 - x_3}\,.
    \end{eqnarray}

    Furthermore, we define the loop function at one-loop level as
    \begin{eqnarray}
      B_{(n_1,\cdots)}(p^2, x_1,\cdots;x_3)&=&(16\pi^2\mu^{2\epsilon})\int \frac{d^dq}{i(2\pi)^{d}}
       \frac{1}{[q^2-x_1]^{n_1}\cdots[(p+q)^2-x_3]}\,.
    \end{eqnarray}

    \subsection{\texorpdfstring{$I_{(2;2)}$}{I{(2;2)}}}
    
    The two-loop function $I_{(2;2)}$ does not contain the UV divergence. 
    $I_{(2;2)}(x_1;x_2;x_3)$ is a symmetric function in variables $x_1$ and $x_2$.
    In the case of $x_1,\,x_2\ll x_3$, 
    $I_{(2;2)}$ is given as 
    \begin{eqnarray}
      I_{(2;2)}(x_1;x_2;x_3)&\simeq &\frac{1}{x_3}\left(\frac{\pi^2}{3}+ \log\frac{x_1x_2}{x_3^2}+
      \log\frac{x_1}{x_3}\log\frac{x_2}{x_3}\right)\,.
    \end{eqnarray}
    In the case of $x_1\ll x_2 < x_3$, we find
    \begin{eqnarray}
      I_{(2;2)}(x_1;x_2;x_3)&\simeq&
      \frac{x_3}{(x_3-x_2)^2}
      \left\{
       \frac{\pi^2}{3}
        -\frac{x_2}{x_3} \log\frac{x_1}{x_2}
        + \log\frac{x_1x_2}{x_3^2}  + \log\frac{x_1}{x_3}\log\frac{x_2}{x_3}\right. \nonumber\\
    &&\qquad \left.
        -2 \left[
        \log\frac{x_2}{x_3}
        \log\left(1-\frac{x_2}{x_3}\right)
        +{\rm Li}_2\left(\frac{x_2}{x_3}\right)
        \right]
        \right\}\,,
    \end{eqnarray}  
    while for $x_1\ll x_3 < x_2$
    \begin{align}
      I_{(2;2)}(x_1;x_2;x_3)\simeq
      \frac{x_3}{(x_3-x_2)^2}
      \left[
        -\frac{x_2}{x_3} \log\frac{x_1}{x_2}
        +\log\frac{x_1x_2}{x_3^2}+  \log\frac{x_1}{x_3}\log\frac{x_2}{x_3} + 2 {\rm Li}_2\left(1-\frac{x_2}{x_3}\right)
        \right]\,.
    \end{align}  

\bibliographystyle{utphys28mod}
\bibliography{ref}

\end{document}